\documentclass[twocolumn,preprintnumbers,amsmath,amssymb,superscriptaddress,nofootinbib,english]{revtex4-1}
\pdfoutput=1
\usepackage{times,amsmath,amsfonts,amssymb,epstopdf}
\usepackage{graphicx}
\usepackage{dcolumn}
\usepackage{bm}
\usepackage{epsfig}
\usepackage{graphicx}
\usepackage{hyperref}
\usepackage[usenames]{color}
\usepackage{url}
\usepackage[normalem]{ulem}
\usepackage[T1]{fontenc}
\usepackage{amsmath}

\usepackage[usenames]{color}

\def\be{\begin{equation}}
\def\ee{\end{equation}}
\def\ben{\begin{eqnarray}}
\def\een{\end{eqnarray}}
\def\ba{\begin{array}}
\def\ea{\end{array}}

\newcommand{\bq}{\begin{eqnarray}}
\newcommand{\eq}{\end{eqnarray}}
\newcommand{\bes}{\begin{subequations}}
\newcommand{\ees}{\end{subequations}}

\begin{document}
\newcommand{\half}{{\textstyle\frac{1}{2}}}
\allowdisplaybreaks[3]
\def\triangledown{\nabla}
\def\grad3{\hat{\nabla}}
\def\a{\alpha}
\def\b{\beta}
\def\g{\gamma}\def\G{\Gamma}
\def\d{\delta}\def\D{\Delta}
\def\ep{\epsilon}
\def\et{\eta}
\def\z{\zeta}
\def\t{\theta}\def\T{\Theta}
\def\l{\lambda}\def\L{\Lambda}
\def\m{\mu}
\def\f{\phi}\def\F{\Phi}
\def\n{\nu}
\def\r{\rho}
\def\s{\sigma}\def\S{\Sigma}
\def\ta{\tau}
\def\x{\chi}
\def\o{\omega}\def\O{\Omega}
\def\k{\kappa}
\def\pa {\partial}
\def\ov{\over}
\def\br{\\}
\def\ud{\underline}

\def\lcdm{\Lambda{\rm CDM}}
\def\qcdm{{\rm QCDM}}
\def\nloc{R\Box^{-2}R}
\def\msun{M_{\odot}/h}
\def\dw{f(X)}

\newcommand\lsim{\mathrel{\rlap{\lower4pt\hbox{\hskip1pt$\sim$}}
    \raise1pt\hbox{$<$}}}
\newcommand\gsim{\mathrel{\rlap{\lower4pt\hbox{\hskip1pt$\sim$}}
    \raise1pt\hbox{$>$}}}
\newcommand\esim{\mathrel{\rlap{\raise2pt\hbox{\hskip0pt$\sim$}}
    \lower1pt\hbox{$-$}}}
\newcommand{\dpar}[2]{\frac{\partial #1}{\partial #2}}
\newcommand{\sdp}[2]{\frac{\partial ^2 #1}{\partial #2 ^2}}
\newcommand{\dtot}[2]{\frac{d #1}{d #2}}
\newcommand{\sdt}[2]{\frac{d ^2 #1}{d #2 ^2}}    

\title{Galaxy cluster lensing masses in modified lensing potentials}

\author{Alexandre Barreira}
\email[Electronic address: ]{a.m.r.barreira@durham.ac.uk}
\affiliation{Institute for Computational Cosmology, Department of Physics, Durham University, Durham DH1 3LE, U.K.}
\affiliation{Institute for Particle Physics Phenomenology, Department of Physics, Durham University, Durham DH1 3LE, U.K.}

\author{Baojiu Li}
\affiliation{Institute for Computational Cosmology, Department of Physics, Durham University, Durham DH1 3LE, U.K.}

\author{Elise Jennings}
\affiliation{Center for Particle Astrophysics, Fermi National Accelerator Laboratory MS209, P.O. Box 500, Kirk Rd. \& Pine St., Batavia, IL 60510-0500}
\affiliation{Kavli Institute for Cosmological Physics, Enrico Fermi Institute, University of Chicago, Chicago, IL 60637}

\author{Julian Merten}
\affiliation{Department of Physics, University of Oxford, Keble Road, Oxford OX1 3RH, U.K.}

\author{Lindsay King}
\affiliation{Department of Physics, The University of Texas at Dallas, Richardson, Texas 75080, USA}

\author{Carlton M. Baugh}
\affiliation{Institute for Computational Cosmology, Department of Physics, Durham University, Durham DH1 3LE, U.K.}

\author{Silvia Pascoli}
\affiliation{Institute for Particle Physics Phenomenology, Department of Physics, Durham University, Durham DH1 3LE, U.K.}

\preprint{IPPP/15/ 26 DCPT/15/ 52}

\begin{abstract}

We determine the concentration-mass relation of 19 X-ray selected galaxy clusters from the CLASH survey in theories of gravity that directly modify the lensing potential. We model the clusters as NFW haloes and fit their lensing signal, in the Cubic Galileon and Nonlocal gravity models, to the lensing convergence profiles of the clusters. We discuss a number of important issues that need to be taken into account, associated with the use of {nonparametric and parametric lensing methods, as well as} assumptions about the background cosmology. Our results show that the concentration and mass estimates in the modified gravity models are, within the errorbars, the same as in $\lcdm$. This result demonstrates that, for the Nonlocal model, the modifications to gravity are too weak at the cluster redshifts, and for the Galileon model, the screening mechanism is very efficient inside the cluster radius. However, at distances $\sim \left[2-20\right] {\rm Mpc}/h$ from the cluster center, we find that the surrounding force profiles are enhanced by $\sim20-40\%$ in the Cubic Galileon model. This has an impact on dynamical mass estimates, which means that tests of gravity based on comparisons between lensing and dynamical masses can  also be applied to the Cubic Galileon model.

\end{abstract} 
%\pacs{98.80.Cq}
\maketitle
%===========================================================================================================================================%
% INTRUDUCTION
%===========================================================================================================================================%
\section{Introduction}\label{sec:into}

Over the past decade, the $\lcdm$ paradigm has established itself as the standard model of cosmology. Most of the matter in this model is in the form of cold dark matter (CDM), and a cosmological constant, $\Lambda$, plays the role of the {\it dark energy} that is responsible for the late-time accelerated expansion of the Universe. Photons, massive neutrinos and baryons make up the rest of the energy budget. The final ingredient is Einstein's theory of General Relativity (GR), which describes the gravitational interaction between these components. Although it is true that this model is in remarkable agreement with most of the cosmological data gathered to date \cite{Planck:2015xua}, it is also the case that, in some aspects, the model still lacks compelling theoretical and observational support. Perhaps the most worrying of these shortcomings is the unnaturally small value of $\Lambda$ compared to the predictions from quantum field theory. Another problem relates to the extrapolation of the regime of validity of GR from the Solar System (where it has been very well tested \cite{Will:2014xja}) to cosmological scales, where there is still a shortage of stringent model-independent tests of gravity. These two problems of $\lcdm$ have been fuelling interest in cosmological models with {\it modified theories of gravity}, which is now a well developed branch of cosmology on both the theoretical \cite{2012PhR...513....1C, Joyce:2014kja} and observational \cite{Jain:2007yk, Huterer:2013xky, Jain:2013wgs, Barreira:2014jha, Koyama:2015vza} levels.

Here, we focus on observational determinations of galaxy cluster masses derived from lensing in theories of modified gravity. This is a topic that has not been extensively investigated in the literature. The reason for this, we believe, is historical as many of the first modified gravity models to be compared to observations were models like $f(R)$ \cite{Sotiriou:2008rp} or Dvali-Gabadadze-Porrati (DGP) \cite{Dvali:2000hr} gravity, which do not modify the lensing potential directly through a modified Poisson equation. In these models, which are conformally equivalent to scalar-tensor theories, lensing mass estimates are automatically the same as in GR, whereas dynamical mass estimates are not \cite{Zhang:2007nk, 2010PhRvD..81j3002S}. This led to the development of a number of tests of gravity. For instance, Refs.~\cite{2010ApJ...708..750S, 2009arXiv0907.4829S} compared galaxy lensing masses from strong lensing with dynamical masses inferred from stellar velocities to probe gravity on ${\rm kpc}$ scales. Also, Refs.\cite{2012PhRvL.109e1301L, Lam:2013kma, Zu:2013joa} developed methods to probe the modified dynamical potential in $f(R)$ and DGP models in the infall regions around massive clusters, given the lensing mass. More recently, Ref.~\cite{Wilcox:2015kna} used comparisons between the X-ray surface brightness and lensing profiles of galaxy clusters to constrain models like $f(R)$ (see also Ref.~\cite{Terukina:2012ji}). The unmodified lensing potential in these theories also allowed for cluster lensing masses to be used as a relatively model independent ingredient in observational tests of gravity. For instance, in the work of Refs.~\cite{2009PhRvD..80h3505S, 2012PhRvD..85l4038L, Cataneo:2014kaa}, the authors used the fact that $f(R)$ models modify the halo mass function to place observational constraints using data from the abundance of clusters as a function of their lensing mass.

Recently, there has been growing interest in models that also modify the way in which the lensing potential depends on matter density perturbations, such as Nonlocal gravity \cite{Deser:2007jk, Deser:2013uya, Maggiore:2014sia, Dirian:2014ara, Dirian:2014bma, Barreira:2014kra}, Galileon gravity \cite{PhysRevD.79.064036, PhysRevD.79.084003, Deffayet:2009mn}, massive gravity \cite{Gabadadze:2009ja, deRham:2009rm, deRham:2010kj, deRham:2010gu, deRham:2010ik, deRham:2010tw,  Hassan:2011zd}, K-mouflage gravity \cite{Brax:2014wla, Brax:2014yla, Barreira:2014gwa, Babichev:2009ee}, and several other subclasses of Horndeski's general theory \cite{Horndeski:1974wa}. These modifications to the lensing signal give rise to a broader range of ways to test gravity. For example, Refs.~\cite{2011PhRvL.106t1102W, Park:2014aga} presented forecasts for future galaxy-galaxy lensing observations \cite{2013MNRAS.432.1544M} showing characteristic signatures of some models of massive gravity. Models that change the lensing signal can also have a strong impact on the power spectrum of cosmic shear \cite{Battye:2014xna, Leonard:2015hha} and the lensing of cosmic microwave background (CMB) photons \cite{Zhao:2008bn, 2012PhRvD..86l4016B, Barreira:2014jha}. Moreover, photons are a direct probe of the time evolution of gravitational potentials, which allows strong constraints to be placed upon modified gravity models via the integrated Sachs-Wolfe (ISW) effect \cite{Barreira:2014jha, Enander:2015vja, 2012PhRvD..85l3503K}. 

{Another consequence of the modifications to the lensing potential is that this may introduce model-dependent systematics in the estimation of cluster masses from lensing. The investigation of such biases and their connection with some of the above-mentioned tests of gravity is one of the main goals of this paper. We choose the Galileon and Nonlocal gravity cosmologies as working examples of models that directly modify the lensing potential.} In the Galileon model, an extra scalar degree of freedom gives rise to a fifth force at late times. The effects of this fifth force are appreciable on large cosmological scales, but are suppresed near massive bodies by means of a screening mechanism known as the {\it Vainshtein effect} {\cite{Vainshtein1972393, Koyama:2013paa, Babichev:2013usa}. This model has been shown to provide a good fit to the CMB temperature, CMB lensing and baryonic acoustic oscillations (BAO) data in Refs.~\cite{Barreira:2014jha, Barreira:2014ija}. Nonlinear structure formation in this model has been studied in Ref.~\cite{Barreira:2013xea} using the spherical collapse model and in Refs.~\cite{Barreira:2013eea, Li:2013tda} using N-body simulation. Reference~\cite{Barreira:2014zza} studied the properties of dark matter haloes, which were used to develop a halo model of structure formation for Galileon gravity. This model is, however, under observational tension as it predicts a negative sign for the ISW effect \cite{Barreira:2014jha}, which is at odds with recent observations \cite{Ade:2015dva}. In the case of the Nonlocal gravity model, the modifications to gravity on cluster scales can be fully parametrized by a time-dependent effective gravitational strength. This model has no screening mechanism, but Refs.~\cite{Kehagias:2014sda, Maggiore:2014sia} have shown that, if the background evolution can be neglected locally, then the model becomes compatible with Solar System tests of gravity. Linear structure formation has been studied by Refs.~\cite{Dirian:2014ara, Dirian:2014bma} and Ref.~\cite{Barreira:2014kra} performed the first N-body simulations of Nonlocal gravity cosmologies, which were used to study halo properties and to also construct a halo model formalism. In addition to the modified gravitational law, both the Galileon and Nonlocal gravity models also modify the expansion rate at late times. This is different from models such as $f(R)$ or DGP gravity for which the expansion rate can be tuned to match that of $\lcdm$.

To estimate lensing masses in Galileon and Nonlocal cosmologies, we model galaxy clusters as Navarro-Frenk-White (NFW) haloes \cite{Navarro:1996gj}, and fit the predicted lensing convergence signal to the data obtained from weak and strong lensing observations for 19 X-ray selected clusters from the Cluster Lensing and  Supernova Survey with the {\it Hubble Space Telescope} (CLASH) \cite{2012ApJS..199...25P, Merten:2014wna, Umetsu:2014vna}. Our analysis is similar to that performed in the context of GR in Ref.~\cite{Merten:2014wna}. In most of the current data analysis, one often makes model-dependent assumptions which may lead to results that are biased towards the assumed models. For example, the analysis of Ref.~\cite{Merten:2014wna} assumes a fiducial $\lcdm$ background to compute angular diameter distances. Assumptions like these must be identified and carefully assessed before using the observations to test alternative models. On the other hand, the lensing data analysis of Ref.~\cite{Merten:2014wna} makes no assumptions about the mass distribution of the clusters, which reduces the model dependency of the conclusions drawn from the observations, and makes it particularly well suited to tests of modified gravity. Given the subtle nature of some steps involved in the analysis of lensing data we shall pay special attention to them and explain how they can be taken into account.

The rest of this paper is organized as follows. In Sec.~\ref{sec:equations}, we describe the calculation of the lensing convergence of NFW haloes in $\lcdm$, Galileon and Nonlocal gravity models, and in Sec.~\ref{sec:fixedparams} we analyse the general predictions for each model. We describe our fitting methodology in Sec.~\ref{sec:clashlike}, where we comment also on the extra steps that one needs to take, in order to account for certain model-dependent assumptions made in the data analysis. In Sec.~\ref{sec:results}, we present our main results for the cluster lensing mass and concentration estimates in the three models we consider. We also discuss the link between the results found here and tests of gravity on large scales, namely those that were first designed for models that do not modify lensing. We summarize in Sec.~\ref{sec:summary}.

\section{Lensing equations}\label{sec:equations}

In this section we specify our notation and describe the calculation of the lensing quantities in the models we consider.
%===========================================================================================================================================%
% LENSING BASICS
%===========================================================================================================================================%
\subsection{Cluster lensing basics}

Throughout, we work under the commonly adopted setup for cluster lensing studies (see e.g.~Refs.~\cite{2010arXiv1002.3952U, Bartelmann:1999yn, 2010CQGra..27w3001B} for comprehensive reviews). In particular,  we consider a set of source galaxies at redshift $z_s$, whose light gets deflected by a galaxy cluster at $z_d$. We use $D_d$, $D_s$ and $D_{ds}$ to denote, respectively, the angular diameter distances between the observer and the lens, the observer and the sources, and the lens and the sources. We assume clusters are spherically symmetric and use the thin-lens approximation in which one neglects the thickness of the galaxy cluster compared to the much larger values of $D_d$, $D_s$ and $D_{ds}$. We also neglect the lensing distortions induced by foreground and background structures, compared to the lensing signal of the cluster. In our notation, $r = \sqrt{x^2 + y^2}$ is a two-dimensional radial coordinate defined on the lens plane and with origin at the cluster center ($x$ and $y$ are cartesian coordinates); $l$ denotes the optical axis (line-of-sight) direction, perpendicular to the lens plane, and with origin also at the cluster center; and $R = \sqrt{r^2 + l^2}$, is a three-dimensional radial coordinate with origin at the cluster center.

%$D_{ds} = D_s - (1+z_d)D_d/(1+z_s)$

Light rays coming from the sources are deflected at the lens position by an angle $\vec{\alpha}$, which is related to the true (unobserved) angular position, $\vec{\beta}$, and the observed one, $\vec{\theta}$, by

\bq\label{eq:lenseq}
\vec{\beta} =\vec{\theta} - \vec{\alpha}(\vec{\theta}).
\eq
The local properties of the lensing signal are fully determined by spatial second derivatives of the scaled projected lensing potential of the galaxy cluster, $\psi$, which is given by 
\bq\label{eq:projpot}
\psi(\theta = r/D_d) = \frac{D_{ds}}{D_dD_s}\frac{2}{c^2}\int_{-D_d}^{D_{ds}} \Phi_{\rm len}(r, l){\rm d}l,
\eq
where $c$ is the speed of light and $\Phi_{\rm len} \equiv \left(\Phi + \Psi\right)/2$ is the total three-dimensional lensing potential. The two Newtonian potentials $\Phi$ and $\Psi$ are defined by the perturbed Friedmann-Robertson-Walker (FRW) line element:
\bq
{\rm d}s^2 = \left(1 + 2\Psi/c^2\right)c^2{\rm d}t^2 - a^2\left(1 - 2\Phi/c^2\right){\rm d}{\bf x}^2,
\eq
where $a = 1/(1+z)$ is the cosmological scale factor ($z$ is the redshift). The Jacobian matrix of the lensing mapping of Eq.~(\ref{eq:lenseq}) is given by
\begin{equation}\everymath{\displaystyle}\frac{\partial \vec{\beta}}{\partial \vec{\theta}}(\vec{\theta})\label{eq:lensing_matrix}
=
\begin{bmatrix}
1 - \kappa - \gamma_1 & -\gamma_2  \\
-\gamma_2 & 1 - \kappa + \gamma_1
\end{bmatrix},
\end{equation}
where
\bq\label{eq:kappa}
\kappa(\theta) &=& \frac{1}{2}\bar{\nabla}^2_{\theta}\psi = \frac{1}{2}\left(\partial^2_{\theta_x} + \partial^2_{\theta_y}\right)\psi, \\ \nonumber
&=&\frac{D_d^2}{2}\bar{\nabla}^2_{r}\psi = \frac{D_d^2}{2}\left(\partial^2_{x} + \partial^2_{y}\right)\psi
\eq
is the {\it lensing convergence}\footnote{The overbar on the $\nabla$ operator indicates that it is the two-dimensional Laplacian. Also, note that $r = D_d\theta$.}, and
\bq\label{eq:shear}
\gamma_1 &=& \frac{1}{2}\left(\partial^2_{\theta_x} - \partial^2_{\theta_y}\right)\psi, \nonumber \\
\gamma_2 &=& \partial_{\theta_x}\partial_{\theta_y}\psi,
\eq
are the two components of the {\it complex lensing shear}, $|\gamma| = \sqrt{\gamma_1^2 + \gamma_2^2}$. The convergence is responsible for an isotropic focusing (or defocusing) of the light rays, whereas the shear field causes distortions in the shapes of the observed source galaxies.

In lensing studies, one can split the analysis into the weak and strong lensing regimes. In the weak lensing regime, the directly observable quantity is the locally averaged complex ellipticity field in the lens plane, $\left<\epsilon\right>$, which can be constructed from measurements of background galaxy shapes. At each point of the lens field, an average is taken over a number of nearby sources to smooth out the intrinsic ellipticity of the galaxies from that caused by the lens (see e.g.~\cite{Bartelmann:1999yn, 2012arXiv1204.4096K, 2013MNRAS.429..661M}). Observationally, the field $\left<\epsilon\right>$ is directly related to the {\it reduced shear}, $g$, (see e.g. Sec.~4 of Ref.~\cite{Bartelmann:1999yn})
\bq\label{eq:redshear}
\left<\epsilon\right> \longleftrightarrow g \equiv \gamma/(1 - \kappa).
\eq
The strong lensing regime takes place in the inner most regions of the lens. There, the lensing quantities $\kappa$ and $\gamma$ become large and the equations become highly nonlinear. As a consequence, highly distorted images like giant arcs or arclets and multiple images of the same background source can form. This happens close to the location of the critical curves of the lens, which are defined as the set of points on the lens plane where the lensing matrix, Eq.~(\ref{eq:lensing_matrix}), becomes singular, i.e., 
\bq\label{eq:detmatrixx}
{\rm det}\left(\partial\vec{\beta}/\partial\vec{\theta}\right) = \left(1 - \kappa\right)^2 - \gamma^2 = 0.
\eq
Observationally, one identifies multiple images and giant-arcs to infer the position and shape of the critical lines. Then, given a theoretical prediction for $\kappa$ and $\gamma$, one can check if ${\rm det}\left(\partial\vec{\beta}/\partial\vec{\theta}\right)$ vanishes at the location of the critical lines. 
%In Sec.~\ref{sec:clashlike}, we shall discuss in more detail the steps involved in the data analysis.

%===========================================================================================================================================%
% LCDM kappa
%===========================================================================================================================================%

\subsection{Convergence in $\lcdm$}

In GR, in the absence of anisotropic stress, $\Phi = \Psi$, and as a result, the lensing potential is equal to the dynamical potential, $\Phi_{\rm len} = \left(\Phi + \Psi\right)/2 = \Phi = \Psi$. Both satisfy the Poisson equation,
\bq\label{eq:poisson_lcdm}
\nabla_{(r,l)}^2\Phi{(r,l)} = 4\pi G \rho(r,l),
\eq
where $G$ is Newton's constant and $\rho(r,l)$ is the three-dimensional density distribution. The lensing convergence is obtained by integrating Eq.~(\ref{eq:poisson_lcdm}) along the line of sight,

\bq\label{eq:intpoissongr}
\int\nabla^2_{(r,l)}\Phi(r,l){\rm d}l = 4\pi G \int\rho(r,l){\rm d}l = 4\pi G \Sigma(r),
\eq
where $\Sigma(r)$ is the surface mass density. The left-hand side of this equation can be manipulated as follows:

\bq\label{eq:manlhs}
&&\int \nabla^2_{(r, l)}\Phi(r,l){\rm d}l = \int \bar{\nabla}^2_{r}\Phi(r,l){\rm d}l + \int \partial^2_l\Phi(r, l){\rm d}l \nonumber \\
&=& \bar{\nabla}^2_{r} \int \Phi(r, l){\rm d}l + \left[ \partial_l\Phi(r, l)\right]^{+\infty}_{-D_d} \approx \frac{D_dD_sc^2}{2D_{ds}}\bar{\nabla}^2_{r} \psi (r)  \nonumber \\
&=& \frac{D_sc^2}{D_{ds}D_d}\kappa(r = \theta D_d),
\eq
where we have used Eqs.~(\ref{eq:projpot}) and (\ref{eq:kappa}), and also the fact that first spatial derivatives of $\Phi$ are negligible at cosmological distances away from the lens (thin-lens approximation). Combining Eqs.~(\ref{eq:intpoissongr}) and (\ref{eq:manlhs}) yields

\bq\label{eq:kappaingr}
\kappa(\theta) = \frac{4\pi G}{c^2}\frac{D_{ds}D_d}{D_s}\Sigma(\theta) \equiv \frac{\Sigma(\theta)}{\Sigma_c},
\eq
where 
\bq\label{eq:sigmacrit}
\Sigma_c = \frac{c^2}{4\pi G}\frac{D_s}{D_{ds}D_d}
\eq
is called the critical surface mass density for lensing. Therefore, in GR, the calculation of the lensing convergence reduces to the evaluation of the projected two-dimensional density profile of the cluster, which can often be done analytically (see Sec.~\ref{subsec:nfw} below). The Hubble expansion rate in $\lcdm$ is given by

\bq\label{eq:Elcdm}
\left(\frac{H(z)}{H_0}\right)^2 = E^2(z) &=&\Omega_{m0}(1+z)^3 + (1 - \Omega_{m0}),
\eq
where $H_0 = 100h\ {\rm km/s/Mpc}$ is the present-day Hubble expansion rate (the subscript "$0$" denotes present-day values) and $\Omega_{m0}$ is the fractional background energy density of pressureless matter. Here, and throughout, we assume a spatially flat Universe and neglect the contribution to the expansion rate from radiation and massive neutrinos. Equation (\ref{eq:Elcdm}) is used in the calculation of the angular diameter distances that enter Eq.~(\ref{eq:sigmacrit}) and the relation between radial and angular scales, $r = D_d\theta$.

%===========================================================================================================================================%
% GALI kappa
%===========================================================================================================================================%

\subsection{Convergence in Galileon Gravity}

We focus on the cubic sector of the Galileon gravity model \cite{PhysRevD.79.064036, PhysRevD.79.084003, Deffayet:2009mn}. Its action is given by
\bq\label{Galileon action}
&& S = \int {\rm d}^4x\sqrt{-g} \left[ \frac{\mathcal{R}}{16\pi G} - \frac{1}{2}c_2\mathcal{L}_2 - \frac{1}{2}c_3\mathcal{L}_3 - \mathcal{L}_m\right], \nonumber \\
\eq
where $\mathcal{R}$ is the Ricci scalar, $g$ is the determinant of the metric $g_{\mu\nu}$, $c_2$ and $c_3$ are dimensionless constants, and $\mathcal{L}_2$ and $\mathcal{L}_3$ are given by
\bq\label{L's}
\mathcal{L}_2 = \nabla_\mu\varphi\nabla^\mu\varphi,\ \ \ \ \ \ \ \  \mathcal{L}_3 = \frac{2}{\mathcal{M}^3}\Box\varphi\nabla_\mu\varphi\nabla^\mu\varphi,
\eq
in which $\varphi$ is the Galileon field, $\mathcal{M}^3 = M_{\rm Pl}H_0^2$, $M_{\rm Pl}^2 = 1/(8 \pi G)$ is the reduced Planck mass, $\Box = \nabla^\mu\nabla_\mu$ is the d'Alembert operator, $\mathcal{L}_m$ is the matter Lagrangian density and Greek indices run over $0$, $1$, $2$, $3$. In flat spacetime, the above action is invariant under the {\it Galilean shift} $\partial_\mu\varphi \rightarrow \partial_\mu\varphi + b_\mu$ (where $b_\mu$ is a constant four-vector). In spherical coordinates, the Poisson equation in the Galileon model leads to the following force law (see Refs.~\cite{Barreira:2013eea, Barreira:2013xea} for details about the derivation)
\bq\label{eq:intpoissongali}
\frac{\Phi,_R}{R} = \frac{GM(<R)}{R^3} - \frac{c_3}{\mathcal{M}^3}\dot{\bar{\varphi}}^2\frac{\delta\varphi,_R}{R},
\eq
where $\delta\varphi(R)$ is the spatial perturbation of the Galileon scalar field about the background value, $\bar{\varphi}(z)$, $M(<R) = 4\pi \int_0^R \rho(r')r'^2{\rm d}r'$ is the mass enclosed inside radius $R$  and $,_R$ denotes partial differentiation w.r.t.~$R$. Equation (\ref{eq:intpoissongali}) differs from GR by having an extra source term which is governed by
\bq\label{eq:galieomsol}
\frac{\delta\varphi,_R}{R} = \frac{4}{3}\frac{M_{\rm Pl}}{\beta_2}\left(\frac{R}{r_V}\right)^3\left[\sqrt{\left(\frac{r_V}{R}\right)^3 + 1} - 1\right]\frac{GM(<R)}{R^3}, \nonumber \\
\eq
with
\bq\label{eq:rv}
r_V^3 = \frac{16}{9}\frac{M_{\rm Pl}}{\beta_1\beta_2\mathcal{M}^3}GM(<r),
\eq
where $\beta_1$ and $\beta_2$ are two dimensionless functions of time. The quantity $r_V$ is a radial scale (often called the Vainshtein radius) that roughly determines the distance from the halo/cluster center within which the modifications to gravity are suppressed. The combination of Eqs.~(\ref{eq:intpoissongali}), (\ref{eq:galieomsol}) and its derivatives leads to 
\begin{widetext}
\bq
\label{eq:phiror}\frac{\Phi,_R}{R} &=& \left\{1 - \frac{4}{3}\frac{c_3}{M_{\rm Pl}\mathcal{M}^3}\frac{\dot{\bar{\varphi}}^2}{\beta_2}\left(\frac{R}{r_V}\right)^3\left[\sqrt{\left(\frac{r_V}{R}\right)^3 + 1} - 1\right]\right\}\frac{GM(<R)}{R^3}, \\
\nonumber \\
\label{eq:phirr}\Phi,_{RR} &=& G\left[\frac{M(<R),_R}{R^2} - \frac{2M(<R)}{R^3}\right] - \frac{3}{4}\frac{c_3\beta_1\dot{\bar{\varphi}}^2}{M_{\rm Pl}^2}\left[\sqrt{\left(\frac{r_V}{R}\right)^3 + 1} - 1 + \frac{3}{2}\frac{(r_V/R)^2}{\sqrt{(r_V/R)^3 + 1}}\left(r_V,_R - \frac{r_V}{R}\right)\right].
\eq
\end{widetext}
In the limit of large $R$, Eq.~(\ref{eq:phiror}) can be written as (note that $r_V^3 \rightarrow {\rm constant}$ as $R$ increases)

\bq\label{eq:linforcegali}
\frac{\Phi,_R}{R} = G_{\rm lin}\frac{M(<R)}{R^3},
\eq
with
\bq\label{eq:geffgali}
G_{\rm lin} = G\left(1 - \frac{2}{3M_{\rm Pl}\mathcal{M}^3}\frac{c_3\dot{\bar{\varphi}}^2}{\beta_2}\right),
\eq
being an effective "linearized" time dependent gravitational strength. On the other hand, when $R$ becomes small, it is straightforward to show that (using, for instance, the NFW expressions shown below)

\bq\label{eq:linforcegali}
\frac{\Phi,_R}{R} \approx \frac{GM(<R)}{R^3}.
\eq
That is, at sufficiently small radii, the force in the Galileon model is approximately the same as in GR, which is a direct consequence of the Vainshtein screening mechanism. 

In the Cubic Galileon model one also has that $\Phi = \Psi$ \cite{Barreira:2013xea}, which implies, like in GR in the absence of anisotropic stress, that the lensing potential is equal to the dynamical potential. We compute the convergence in the Cubic Galileon model by numerically integrating the three-dimensional Laplacian of the total potential, $\nabla^2\Phi$, along the line of sight as

\bq\label{eq:kappaingali}
\kappa(\theta) &=& \frac{D_{ds}D_d}{D_sc^2} \int_{-D_d}^{\infty} \nabla^2_{(r, l)}\Phi(r, l) {\rm d}l \nonumber \\
&=& \frac{1}{4\pi G\Sigma_c} \int_{-D_d}^{\infty}\left(\Phi,_{RR}(r, l) + 2\frac{\Phi,_R}{R}(r, l)\right) {\rm d}l \nonumber \\
\eq
(recall that in spherical coordinates, $\nabla^2\cdot \rightarrow \cdot,_{RR} + 2\frac{\cdot,_R}{R}$). 

All that is left to specify is the time dependence of the background quantities that enter the above equations of the Galileon model. The time evolution of the Hubble parameter, $\dot{\bar{\varphi}}$, $\beta_1$ and $\beta_2$ are given, respectively, by \cite{Barreira:2014jha}

\bq\label{eq:galibg}
E(a)^2 &=& \frac{1}{2}\left[\Omega_{m0}a^{-3} + \sqrt{\Omega_{m0}^2a^{-6}  + 4(1 - \Omega_{m0})}\right], \\
\dot{\bar{\varphi}} &=& \xi H_0/E(a),  \\
\beta_1 &=& \frac{1}{6c_3}\left[-c_2 - \frac{4c_3}{\mathcal{M}^3}\left(\ddot{\bar{\varphi}} + 2H\dot{\bar{\varphi}}\right) + \frac{2 c_3^2}{M_{\rm Pl}^2\mathcal{M}^6}\dot{\bar{\varphi}}^4\right], \\
\beta_2 &=& \frac{2\mathcal{M}^3M_{\rm Pl}}{\dot{\bar{\varphi}}^2}\beta_1.
\eq
As in Ref.~\cite{Barreira:2014jha}, we take $c_2 = -1$ and the other two Galileon parameters are determined by $\Omega_{m0}$ as

\bq\label{eq:c3ksi}
\xi &=& \sqrt{6(1-\Omega_{m0})}, \\
c_3 &=&  1/(6\xi).
\eq

%The above background expressions for the Galileon model are obtained by assuming that the Galileon field is evolving according to the so-called {\it tracker} solution, which has been shown to be an observational requisite determined by the CMB data \cite{Barreira:2014jha}. 

%===========================================================================================================================================%
% NLOC kappa
%===========================================================================================================================================%

\subsection{Convergence in Nonlocal Gravity}\label{sec:kappanoloc}

We take the model of Refs.~\cite{Maggiore:2014sia, Dirian:2014ara} as the representative case of a Nonlocal gravity model. The action is given by 
\bq\label{eq:action}
A = \frac{1}{16 \pi G}\int {\rm d}x^4\sqrt{-g}\left[\mathcal{R} - \frac{m^2}{6}\mathcal{R}\Box^{-2}\mathcal{R} - \mathcal{L}_m\right].
\eq
This (nonlocal) action can be cast in a more familiar (local) form given by \cite{Nojiri:2007uq, Capozziello:2008gu, Koshelev:2008ie}
\bq\label{eq:action-local}
A &&= \frac{1}{16 \pi G}\int {\rm d}x^4\sqrt{-g}\left[\mathcal{R} - \frac{m^2}{6}\mathcal{R}S - \xi_1\left(\Box U + \mathcal{R}\right) \right. \nonumber \\
&&\ \ \ \ \ \ \ \ \ \ \ \ \  \ \ \ \ \ \ \ \ \ \ \ \ \ \ \ \ \ \ \ \left. - \xi_2\left(\Box S + U\right) - \mathcal{L}_m\right],
\eq
where $\xi_1$ and $\xi_2$ are two Lagrange multipliers and we have introduced two auxiliary scalar fields, $U = - \Box^{-1} \mathcal{R}$ and $S = \Box^{-2}\mathcal{R}$. For completeness, we note that these two formulations are not equivalent and that care must be taken when matching the solutions associated with the above two actions (see e.g.~Refs.~\cite{Koshelev:2008ie, Koivisto:2009jn, Barvinsky:2011rk, Deser:2013uya, Maggiore:2013mea, Foffa:2013sma, Foffa:2013vma} for a discussion)

Following Refs.~\cite{Dirian:2014ara, Barreira:2014kra}, for the scales relevant for large scale structure formation, the modifed Poisson equation in this Nonlocal model can be written as
\bq\label{eq:modpoisson}
\nabla^2_{(r,l)}\Phi = 4\pi G_{\rm eff}(z) \rho(r,l),
\eq
which takes the same form as in GR, Eq.~(\ref{eq:poisson_lcdm}), but with an effective time dependent gravitational strength given by
\bq
\label{eq:geff}G_{\rm lin} = G \left[1 - \frac{m^2\bar{S}(z)}{3}\right]^{-1},
\eq
where $\bar{S}$ is the background part of the field $S$. The time evolution of the background quantities in the Nonlocal model has to be obtained by numerically integrating the background differential equations (see e.g.~Refs~\cite{Dirian:2014ara, Barreira:2014kra}). The parameter $m$ in Eqs.~(\ref{eq:action}) and (\ref{eq:action-local}) controls the amount of dark energy in the Universe. In a flat Universe, the value of $m$ is therefore determined by the energy densities of the remaining matter species, which means this Nonlocal gravity model has the same number of free parameters as $\lcdm$. 

Just as in the cases of $\lcdm$ and Cubic Galileon gravity, in the Nonlocal model we also have that $\Phi = \Psi = \Phi_{\rm len}$, in the absence of anisotropic stress. Moreover, since the Poisson equation in this model is obtained from GR by a simple rescaling of the gravitational strength, it follows that the convergence can also be computed analytically as

\bq\label{eq:kappainnloc}
\kappa(\theta) = \left(\frac{G_{\rm lin}}{G}\right) \frac{\Sigma(\theta)}{\Sigma_c}.
\eq
Note, however, that the Nonlocal expansion rate must be used in the calculation of the angular diameter distances that enter $\Sigma_c$.

Contrary to the case of Galileon gravity, the Nonlocal model does not possess a screening mechanism, which may raise some concerns about the ability of this model to pass Solar System constraints \cite{Will:2014xja}. For instance, Ref.~\cite{Barreira:2014kra} showed that the rate of change of the gravitational strength on cosmological scales, $\dot{G}_{\rm lin}$, if applied directly to Solar System tests, results in the model becoming inconsistent with current lunar laser ranging experiments \cite{Williams:2004qba}. However, the time variation of $G_{\rm lin}$ follows from the background expansion rate, and it is not clear what its impact is in the Solar System. In fact, the authors of Refs.~\cite{Kehagias:2014sda, Maggiore:2014sia} have shown that if the cosmological expansion is neglected, i.e. if the spacetime about the Sun is perturbed Minkowskii (as opposed to FRW), then the model predictions become compatible with the  current bounds. Here, we shall bear these discussions in mind, but focus instead on the model predictions for cluster scales, which are sufficiently large for one to consider the gravitational strength given by  Eq.~(\ref{eq:geff}).

%===========================================================================================================================================%
% NFW EQUATIONS
%===========================================================================================================================================%

\subsection{NFW halo expressions}\label{subsec:nfw}

In order to compute the lensing convergence in any of the cosmological models considered above, we need to specify the density profile of the lenses, which we model as dark matter haloes with NFW density profiles \cite{Navarro:1996gj},

\bq\label{eq:nfwdensity}
\rho_{\rm NFW}(R, z) = \frac{\rho_s(z)}{(R/r_s)(1 + R/r_s)^2}.
\eq
This profile is fully specified by two parameters known as the {\it scale radius}, $r_s$, and the {\it characteristic density}, $\rho_s$. The mass enclosed inside radius $R$ in a NFW halo is given by:

\bq\label{eq:minr}
M_{\rm NFW}(<R) &=& 4\pi \rho_s r_s^3 \left[{\rm ln}\left(1 + R/r_s\right) - \frac{R/r_s}{1 + R/r_s}\right]. \nonumber \\
\eq
We define halo masses in the usual way

\bq\label{eq:Mdelta}
M_{\Delta} &=& \frac{4\pi}{3}\Delta{\rho}_{c}R_{\Delta}^3,
\eq
where $R_{\Delta}$ is the radius within which the mean density is equal to $\Delta$ times the critical density of the Universe at a given redshift, $\rho_c = 3H^2(z)/(8\pi G)$. Equating $M_{\Delta} = M_{\rm NFW}(<R_{\Delta})$ one finds

\bq\label{eq:rhos}
\rho_s = \frac{1}{3}\Delta {\rho}_{c}c_\Delta^3\left[{\rm ln}\left(1 + c_\Delta\right) - \frac{c_{\Delta}}{1 + c_{\Delta}}\right]^{-1},
\eq
where we have defined the concentration parameter

\bq\label{eq:c}
c_\Delta= R_\Delta/r_s.
\eq
We take $\Delta = 200$ (as it is standard in the literature since it is close to the overdensity at virial equilibrium) and instead of characterizing the NFW haloes by $\rho_s$ and $r_s$, we use the equivalent and more common parametrization in terms of $M_{200}$ and $c_{200}$.

The surface mass density of a NFW halo admits an analytical solution given by \cite{Bartelmann:1996hq, 1999astro.ph..8213O, 2010arXiv1002.3952U}
%\bq\label{eq:sigmanfw}
%&&\Sigma_{\rm NFW}(r = D_d\theta) = \int \rho_{\rm NFW}(r, l){\rm d}l \nonumber \\
%&&= \frac{2r_s\rho_s}{x^2-1} \left(1 - \frac{2}{\sqrt{1 - x^2}}{\rm arctanh}\left[\sqrt{\frac{1-x}{1+x}}\right]\right),\ \ \ {\rm if}\ \ \ x < 1 \nonumber \\
%&&= \frac{2r_s\rho_s}{3},\ \ \ {\rm if}\ \ \ x = 1 \nonumber \\
%&&= \frac{2r_s\rho_s}{x^2-1} \left(1 - \frac{2}{\sqrt{x^2 - 1}}{\rm arctan}\left[\sqrt{\frac{x-1}{1+x}}\right]\right),\ \ \ {\rm if}\ \ \ x > 1, \nonumber \\
%\eq

\bq\label{eq:sigmanfw}
\Sigma_{\rm NFW}(r = D_d\theta) = \int \rho_{\rm NFW}(r, l){\rm d}l = 
\eq
\[ =\begin{cases} 
       \frac{2r_s\rho_s}{x^2-1} \left(1 - \frac{2}{\sqrt{1 - x^2}}{\rm arctanh}\left[\sqrt{\frac{1-x}{1+x}}\ \right]\right)  & x < 1 \\
       \frac{2r_s\rho_s}{3} & x =1  \\
       \frac{2r_s\rho_s}{x^2-1} \left(1 - \frac{2}{\sqrt{x^2 - 1}}{\rm arctan}\left[\sqrt{\frac{x-1}{1+x}}\ \right]\right) & x > 1 
   \end{cases}
\]
where $x = r/r_s$.  The calculation of the lensing convergence in Galileon gravity also requires the evaluation of the gradient of $M_{\rm NFW}(< R)$, which is given by
\bq\label{eq:minrr}
M_{\rm NFW}(<R),_R = 4\pi \rho_s r_s^3\frac{R}{(r_s + R)^2}.
\eq

%===========================================================================================================================================%
% FIXED PARAMS
%===========================================================================================================================================%

\section{Model predictions for fixed cluster parameters}\label{sec:fixedparams}

\begin{figure}
	\centering
	\includegraphics[scale=0.345]{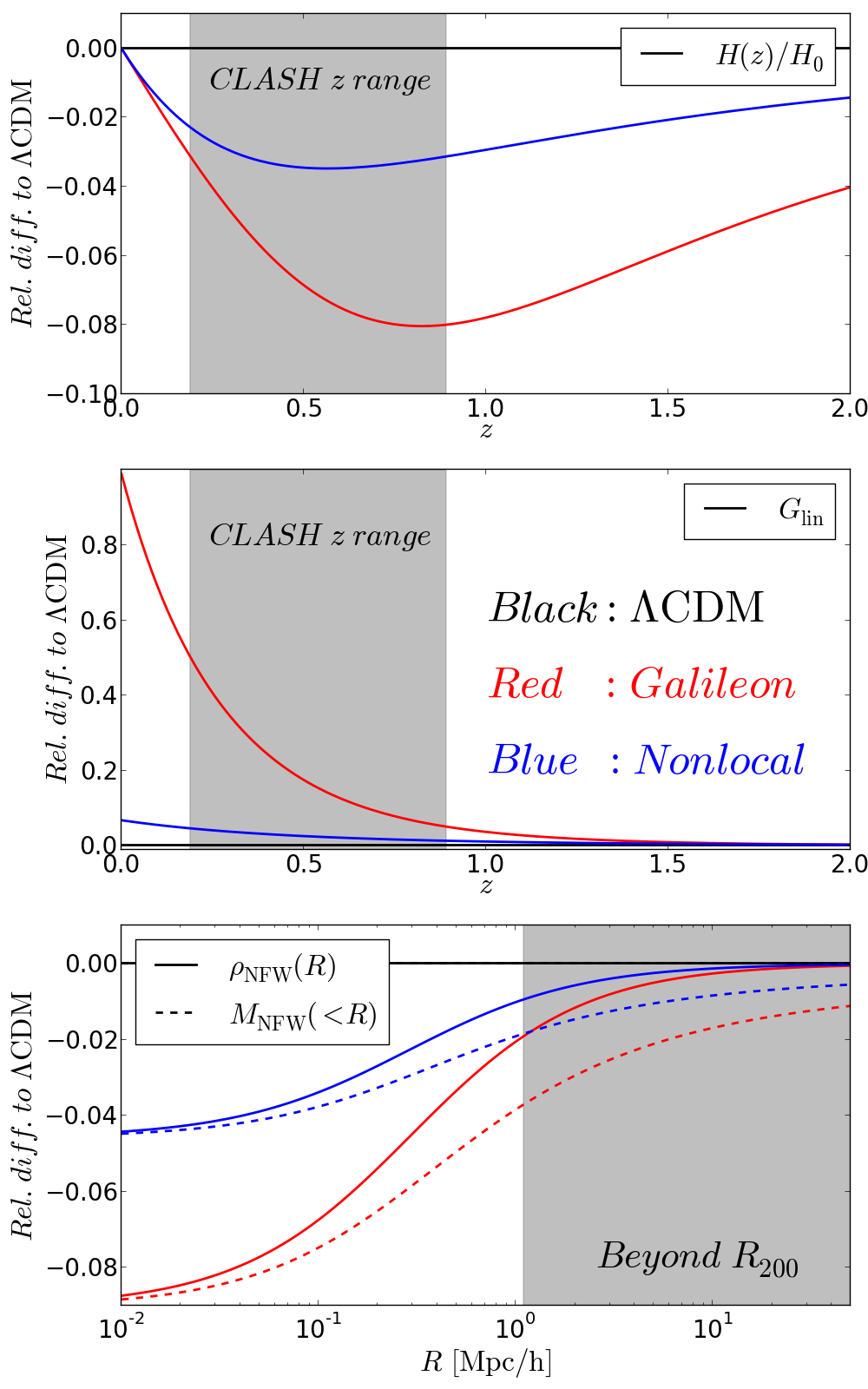}
	\caption{The upper and middle panels show, respectively, the time evolution of the Hubble expansion rate, $H(z)/H_0$, and of the effective linear gravitational strength, $G_{\rm lin}$, for the Galileon (red) and Nonlocal (blue) models, plotted as the relative difference to $\lcdm$. The shaded band in these two panels indicates the redshift range spanned by the CLASH clusters analysed in this paper. The lower panel shows the density (solid) and enclosed mass (dashed) profiles for the Galileon (red) and Nonlocal (blue) models, plotted as the relative difference to $\lcdm$. The NFW parameters and redshift are $M_{200} = 5.0\times10^{14} M_{\odot}/h$, $c_{200} = 4.0$, $z_d = 0.5$. To guide the eye, the shaded band in the lower panel indicates the radial scales outside $R_{200}$ in $\lcdm$.}
\label{fig:bg}\end{figure}

\begin{figure*}
	\centering
	\includegraphics[scale=0.39]{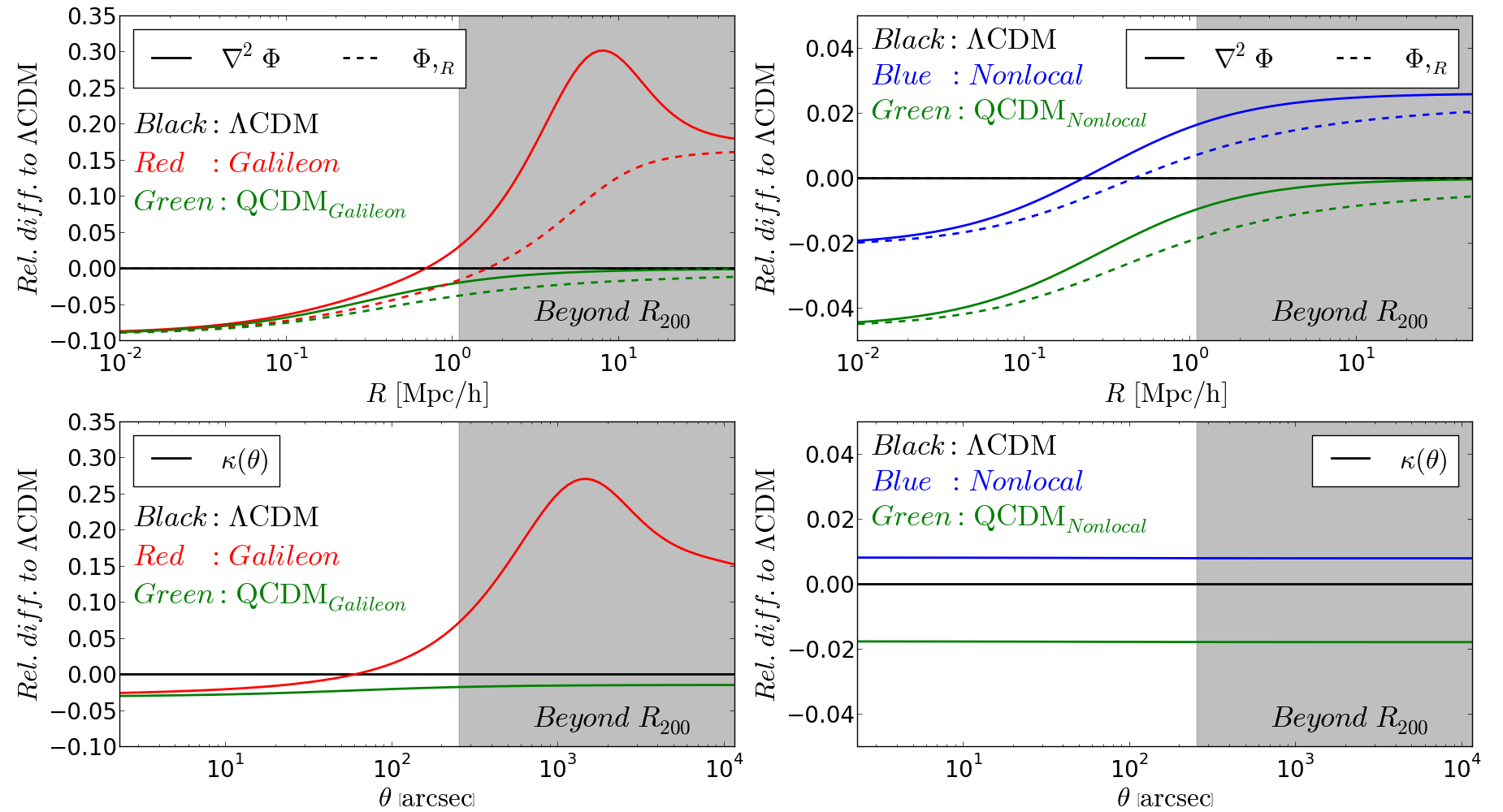}
	\caption{(Upper left) Radial profiles of the Laplacian ($\nabla^2\Phi$, solid) and gradient ($\Phi,_r$, dashed) of the total gravitational potential, $\Phi$, for the Galileon model (red), plotted as the relative difference to $\lcdm$ (black). The NFW parameters and redshift are $M_{200} = 5.0\times 10^{14} M_{\odot}/h$, $c_{200} = 4.0$, $z_d = 0.5$. Also shown are the predictions for a model called ${\rm QCDM}$ (green), which has the same background cosmology as the Galileon model, but with the force calculated as in GR. To guide the eye, the shaded band indicates the radial scales outside $R_{200}$ in $\lcdm$. (Upper right) Same as the upper left panel but for the Nonlocal gravity model (blue). (Lower panels) Same as the upper panels but for the lensing convergence, $\kappa(\theta)$, and assuming $z_s = 1$. In the lower panels we show $\theta$ on the x-axis, which, in the different models, relates differently to physical distances due to the modifications to $D_d$.}
\label{fig:forces}\end{figure*}

To develop intuition about our results, we first analyse the model predictions for fixed cluster parameters, $M_{200}$, $c_{200}$, and cosmological matter density $\Omega_{m0}$. Figure \ref{fig:bg} shows the time evolution of $H(z)/H_0$ (top panel) and $G_{\rm lin}$ (middle panel) for the Galileon and Nonlocal models, plotted as the difference relative to $\lcdm$. The lower panel of Fig.~\ref{fig:bg} shows the NFW density (solid) and enclosed mass (dashed) profiles for a halo at $z_d = 0.5$, and with $M_{200} = 5.0\times 10^{14} M_{\odot}/h$ and $c_{200} = 4$. For all models, the cosmological matter density is $\Omega_{m0} = 0.27$. We note, for completeness, that if $h$ is absorbed into unit definitions (e.g. $M_{\odot}/h$ for masses or ${{\rm Mpc}/h}$ for distances), then our analysis becomes completely independent of its value, for all models.

Figure \ref{fig:bg} shows that the amplitude of the halo density profiles near their center is lower in the Galileon and Nonlocal models, than it is in $\lcdm$. This is because of the lower values of  $\rho_c(z_d)$ in these models. Specifically, in the inner-most part, the density profile of the halo becomes approximately
\bq\label{eq:nfwinner}
\rho_{\rm NFW}(R) \approx \frac{\rho_sr_s}{R} \propto \frac{\rho_c(z_d)^{2/3}}{R},
\eq
as can be checked by noting that $\rho_s \propto \rho_c$ and $r_s \propto \rho_c^{-1/3}$, and recalling that we are assuming fixed $M_{200}$ and $c_{200}$. As a result, if $H(z)$ is smaller than in $\lcdm$ at $z_d$ (upper panels), then so is $\rho_{c}^{2/3}(z_d) \propto H^{4/3}(z_d)$, which effectively leads to a less dense halo. The same qualitative reasoning also applies to the regime where the NFW density scales as $\propto R^{-2}$. However, far from the halo centre, we have
\bq\label{eq:nfwinner}
\rho_{\rm NFW}(R) \approx \frac{\rho_sr_s^3}{R^3} \propto \frac{1}{R^3}.
\eq
In this case, the dependence on $\rho_{c}(z_d)$ cancels out, and hence, all models have the same density values, as seen in the lower panel of Fig.~\ref{fig:bg} ($R \gtrsim 10{\rm Mpc}/h$). The enclosed mass profiles show a qualitatively similar trend. Perhaps the only noteworthy difference is that the $M(<R)$ profiles for the different models do not agree at large radii. In this limit, the enclosed mass (Eq.~(\ref{eq:minr})) scales with radius as
\bq\label{eq:minrhigh}
M(< R) \propto \left[{\rm ln}\left(1 + R/r_s\right) - 1\right],
\eq
which retains a cosmological dependence ($1/r_s \propto \rho_c^{1/3}$) in the logarithmic divergence of the mass. In particular, in the Galileon and Nonlocal models, $\rho_c$ is smaller than in $\lcdm$, which implies that $M(<R)$ is also smaller, as shown.

Another important consequence of the modified Hubble expansion rates in the Galileon and Nonlocal models is in the calculation of the angular diameter distances that determine $\Sigma_c$ and the relation between radial and angular scales. In particular, the smaller values of $H(z)/H_0$ in the Galileon and Nonlocal models lead to larger angular diameter distances, since these are $\propto \int 1/H(z){\rm d}z$.

Figure \ref{fig:forces} shows, for the same parameters as in Fig.~\ref{fig:bg}, the radial profiles of $\nabla^2\Phi$ (solid) and $\Phi,_R$ (dashed) for the Galileon (upper left, red) and Nonlocal (upper right, blue) models. For both models, the figure also shows the predictions from models called ${\rm QCDM}$ (green), which are illustrative models with the same background cosmology as their respective modified gravity models, but keeping the gravitational law of GR. Comparing the ${\rm QCDM}$ variants with the respective {\it full} models allows one to isolate the effects of the modified background from those due to the modified gravitational law. Since both {\rm QCDM} models differ from $\lcdm$ only via the modified background, their relative differences in the profiles of $\nabla^2\Phi \sim \rho$ and $\Phi,_R \sim M(<R)$ are determined only by the background dependence of the density and mass profiles of the halos. As a result, the {\rm QCDM} curves in Fig.\ref{fig:forces} follow the same behavior seen in the lower panel of Fig.~\ref{fig:bg}. It is therefore more interesting to analyse the impact of the fifth forces. In the case of the Nonlocal model, we have seen that the modifications to gravity can be parametrized by a scale-independent, but time-evolving effective gravitational strength, $G_{\rm lin}(z)$, according to Eq.~(\ref{eq:modpoisson}). Consequently, the effect of the fifth force in this model is to boost the amplitude of $\nabla^2\Phi$ and $\Phi,_R$ by the same amount on all scales. From Fig.~\ref{fig:bg}, at $z_d = 0.5$ one has $G_{\rm lin} \approx 0.022$, which corresponds to the difference between the predictions of the ${\rm QCDM}$ and Nonlocal models seen in Fig.~\ref{fig:forces}.

The effects of the fifth force in the Galileon model are slightly more complex due to the nonlinear screening mechanism. By comparing the predictions of the full Galileon model with those of the corresponding ${\rm QCDM}$ variant, one can identify three regimes. The first is a "fully-screened" regime, $R \lesssim 0.1 {\rm Mpc}/h$, where the effects of the fifth force are almost negligible ($\Phi \approx \Phi^{\rm GR}$), as seen by the overlap between the red and green sets of curves. The second regime is a "partly screened" regime which occurs on scales $0.1 {\rm Mpc}/h \lesssim R \lesssim 50 {\rm Mpc}/h$. On these scales, we can write $\Phi = \alpha(r) \Phi_{\rm GR}$, where the function $\alpha$ encapsulates the scale-dependence of the fifth force. Finally, on scales $r \gtrsim 50 {\rm Mpc}/h$, the fifth force becomes completely unscreened and one effectively has $\Phi = G_{\rm lin}(z) \Phi_{\rm GR}$, where $G_{\rm lin}$ is given by Eq.~(\ref{eq:geffgali}) for the Galileon model. This translates into a constant boost in the values of $\nabla^2\Phi$ and $\Phi,_R$ at large radii. The size of this boost in the unscreened regime of the Galileon model ($15-20\%$) is larger than that in the Nonlocal gravity model ($2-3\%$). This follows from the higher value of $G_{\rm lin}$ in the Galileon model at $z_d = 0.5$, as shown in the middle planel of Fig.~\ref{fig:bg}.

The lower panels of Fig.~\ref{fig:forces} show the lensing convergence angular profiles for the Galileon (red) and Nonlocal (blue) models, as well as their respective ${\rm QCDM}$ (green) variants, plotted as the difference relative to $\lcdm$ and assuming $z_s = 1$. These convergence angular profiles relate to the radial profiles of $\nabla^2\Phi$ by (i) the integration along the line of sight; (ii) the overall amplitude scaling set by $\Sigma_c$; (iii) and also importantly, horizontal shifts caused by the fact that the same angular scales correspond to different distance scales at the cluster position because of the different $D_d$ values. In the Galileon and Nonlocal model backgrounds, $D_d$ becomes larger, and as a result, the same radial scales correspond to smaller angular scales. The net result of these effects is to reduce slightly the relative differences of $\kappa$ in the Galileon model w.r.t.~$\lcdm$, compared to the relative differences observed in $\nabla^2\Phi$. The same holds for the case of the Nonlocal model, for which the relative difference becomes also very weakly dependent on the angular scale (the slope of the curves is hardly noticeable in the scale of the figure).

%===========================================================================================================================================%
% CLASH LIKELIHOOD
%===========================================================================================================================================%

\section{Methodology}\label{sec:clashlike}

We estimate cluster masses in Cubic Galileon and Nonlocal gravity cosmologies using the radially-binned lensing convergence profiles obtained from the reconstructions of the lensing potential for 19 X-ray selected galaxy clusters from CLASH \cite{2012ApJS..199...25P}. In this section, we describe our methodology, paying particular attention to a number of subtleties that need to be accounted for to self-consistently compare the data with predictions from the alternative models studied here.

\bigskip

\subsection{Cluster convergence profiles in alternative models}\label{sec:method}

\begin{figure}
	\centering
	\includegraphics[scale=0.345]{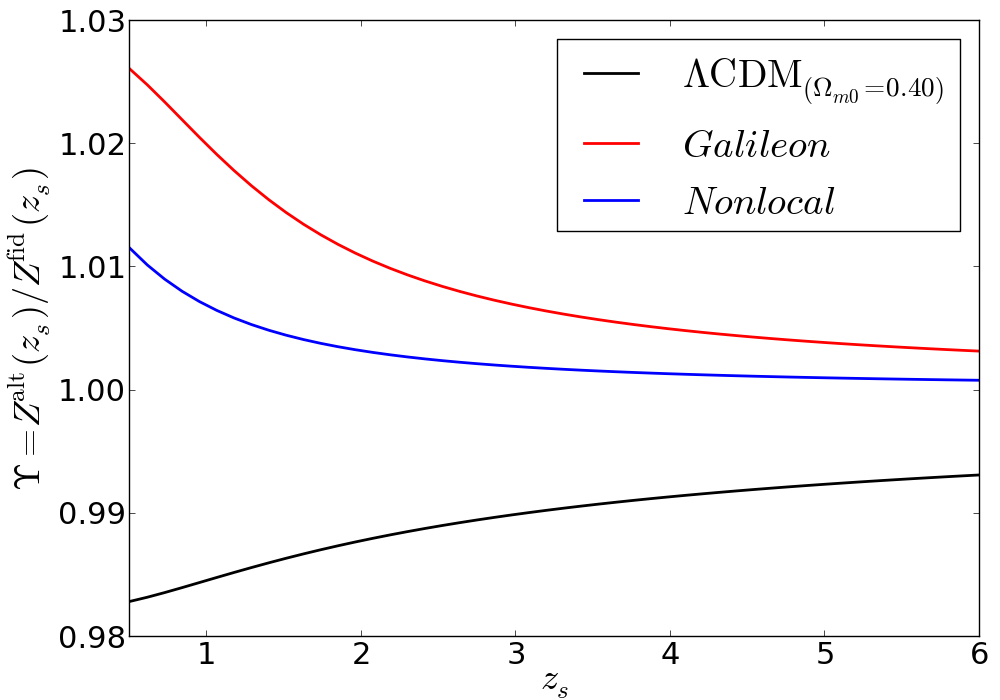}
	\caption{Dependence of the factor $\Upsilon = Z^{\rm alt}(z_d, z_s)/Z^{\rm fid}(z_d, z_s)$, Eq.~(\ref{eq:factor}), on the source redshift $z_s$ for a $\lcdm$ model with $\Omega_{m0} = 0.4$ (black), Galileon gravity (red) and Nonlocal gravity (blue). The cosmological background of these models is different from the fiducial $\lcdm$ model with $\Omega_{m0} = 0.27$ used by Ref.~\cite{Merten:2014wna}. In this figure, $z_d=0.35$, which is typical for the CLASH clusters.}
\label{fig:z}\end{figure}

We use the convergence profile data that was obtained for the CLASH clusters in Ref.~\cite{Merten:2014wna}. There, the analysis was performed with a numerical algorithm called {\tt SaWLens} \cite{Merten:2008qf}, which iteratively reconstructs the lensing potential for each cluster on a two-dimensional grid that covers the cluster field. The analysis is non-parametric, i.e., it makes no assumptions about the mass distribution of the cluster. We refer the reader to Refs.~\cite{Merten:2008qf, Merten:2014wna, 2011MNRAS.417..333M} for the details about how {\tt SaWLens} operates. For the discussion here, what is important to note is that what {\tt SawLens} actually reconstructs is the lensing potential scaled to a source redshift of infinity, $\psi_{\infty} = \psi(z_s = \infty)$, by assuming a fiducial cosmological model. We use $\kappa_{\infty}$ to denote the lensing convergence associated with $\psi_{\infty}$, which is related to the convergence at the true source redshift, $z_s$, via
\bq\label{eq:inftyscale}
\kappa_{z_s} &=& Z^{\rm fid}(z_d, z_s) \kappa_{\infty},
\eq
where we use the subscript $_{z_s}$ to emphasize that $\kappa_{z_s}$ corresponds to the convergence associated with $z_s$. The function $Z \equiv Z(z_d, z_s)$ {\it transports} the convergence from a source redshift of infinity to the source redshift that corresponds to the galaxies on each {\tt SaWLens} grid cell/pixel (we use the words cell and pixel interchangeably). It is given by
\bq\label{eq:Z}
Z^{\rm fid}(z_d, z_s) = \frac{D^{\rm fid}_{s,\infty}D^{\rm fid}_{ds}}{D^{\rm fid}_{ds,\infty}D^{\rm fid}_{s}},
\eq
where the superscript $^{\rm fid}$ indicates angular diameter distances that are calculated assuming the fiducial background cosmology and the subscript $_\infty$ means that the calculation assumes that $z_s = \infty$. In the reconstruction process of Ref.~\cite{Merten:2014wna}, the fiducial cosmology is a $\lcdm$ model with $\Omega_{m0} = 0.27$. From hereon, we use $\kappa^{\rm fid}_{\infty}$ to denote the convergence profiles obtained in this way, where the superscript $^{\rm fid}$ makes it explicit that the data is linked to the fiducial model. It is therefore important to investigate the extent to which the $\kappa^{\rm fid}_{\infty}$ profiles can be used in studies of alternative cosmologies. 

Consider the case that we wish to estimate the lensing masses of the CLASH clusters in a model with a cosmological background that is different from the fiducial model originally used to analyse the observations in Ref.~\cite{Merten:2014wna}. In principle, we could suitably modify the {\tt SaWLens} algorithm to reconstruct the convergence maps in the alternative model, $\kappa^{\rm alt}_{\infty}$, instead of $\kappa^{\rm fid}_{\infty}$. However, this would not be practical as it would imply rerunning the entire analysis pipeline for different background cosmologies. A more economical strategy is to note that the two convergence maps, $\kappa^{\rm fid}_{\infty}$ and $\kappa^{\rm alt}_{\infty}$, can be related by

\bq\label{eq:kapparelation}
Z^{\rm fid}(z_d, z_s)\kappa^{\rm fid}_{\infty} = Z^{\rm alt}(z_d, z_s)\kappa^{\rm alt}_{\infty},
\eq
where $Z^{\rm alt}$ is defined as in Eq.~(\ref{eq:Z}) but with the distances calculated in any alternative, and not the fiducial, cosmology. The above equation holds (up to a correction that we discuss in the next subsection) since both $Z^{\rm fid}\kappa^{\rm fid}_{\infty}$ and $Z^{\rm alt}\kappa^{\rm alt}_{\infty}$ correspond to $\kappa_{z_s}$, i.e., the convergence at the true source redshift \footnote{For example, for $\lcdm$ with an alternative background and for fixed surface mass density, for simplicity, Eq.~(\ref{eq:kapparelation}) becomes $\Sigma(\theta)/\Sigma_c^{\rm alt} =\Sigma(\theta)/\Sigma_c^{\rm fid} \longrightarrow \Sigma_c^{\rm alt} = \Sigma_c^{\rm fid}$.}. Using the above equation, the radially binned convergence profiles obtained using the fiducial cosmology in Ref.~\cite{Merten:2014wna}, $\kappa^{\rm fid}_{\infty}(\theta)$, can be directly compared to the prediction of the alternative model $\kappa^{\rm alt}_{\infty}$, provided the latter is multiplied by the factor $Z^{\rm alt}(z_d, z_s)/Z^{\rm fid}(z_d, z_s)$. This is the approach that we adopt in this paper. Specifically, we aim to obtain constraints on $M_{200}$, $c_{200}$ and $\Omega_{m0}$ in non-fiducial backgrounds by minimizing the $\chi^2$ quantity
\bq\label{eq:chi2nfw}
\chi^2 &=& \overrightarrow{V}\mathcal{C}_{\kappa}^{-1}\overrightarrow{V},
\eq
where
\bq\label{eq:vector}
\vec{V}_i &=& \kappa^{\rm fid}_{\infty, i} - \Upsilon\kappa_{\infty}^{\rm alt}(M_{200}, c_{200}, \Omega_{m0}, \theta_i),
\eq
is the $i$-th entry of the vector $\vec{V}$; $\kappa^{\rm fid}_{\infty, i}$ is the reconstructed lensing convergence in the $i$-th radial bin, $\theta_i$; $\mathcal{C}_{\kappa}$ is the covariance matrix of the radially binned data \footnote{{The bootstrap realizations used to derive the covariance matrices in Ref.~\cite{Merten:2014wna} also make use of the fiducial cosmological background. Here, we use the errors as obtained for the fiducial cosmology and do not attempt to estimate the dependence of the covariance matrix on the assumed cosmology. This does not alter our conclusions as this choice only affects the precise size of the confidence intervals, without introducing any important systematics.}}; and for brevity of notation, we introduce the {\it scaling factor}
\bq\label{eq:factor}
\Upsilon(z_d, z_s) = \frac{Z^{\rm alt}(z_d, z_s)}{Z^{\rm fid}(z_d, z_s)}.
\eq
In Eq.~(\ref{eq:vector}), $\kappa_{\infty}^{\rm alt}$ is given by Eq.~(\ref{eq:kappaingr}) for $\lcdm$, Eq.~(\ref{eq:kappaingali}) for Galileon gravity and Eq.~(\ref{eq:kappainnloc}) for Nonlocal gravity, but using $z_s = \infty$, in the calculation of $\Sigma_c$.

Unless otherwise specified, we assume flat priors on the free parameters, $\Omega_{m0} \in \left[0.1, 0.5\right]$, $M_{200} \in \left[0.3, 3.0\right] \times 10^{15} M_{\odot}/h $ and $c_{200} \in \left[1, 8\right]$.

\subsection{The validity of Eq.~(\ref{eq:kapparelation}) and the choice of source redshifts}\label{sec:validity}

As discussed above, $\kappa^{\rm fid}_{\infty}$ is reconstructed by applying the transformation of Eq.~(\ref{eq:inftyscale}) in each cell of the {\tt SaWLens} grid that covers the cluster field. In this process, the value of $z_s$ is determined by the redshift of the galaxies used to measure the ellipticity field at that pixel, or by the redshift of the galaxies associated with a given multiple image system. On the other hand, our methodology is based on Eq.~(\ref{eq:kapparelation}), in which one scales the lensing quantities from $z_s = \infty$ to a source redshift $z_s$, but neglects the redshift distribution of the background lensed galaxies. The validity of Eq.~(\ref{eq:kapparelation}) then becomes linked to the impact of the spread of the redshift distribution of the source galaxies across each cluster field. For the CLASH clusters analysed in Ref.~\cite{Merten:2014wna}, the redshift distribution of the background galaxies is manifest in four main aspects:

(i) in the weak lensing regime, different ellipticity pixels are associated with different source redshifts since the shapes are measured using different galaxies across the cluster field;

(ii) related to the above, the ellipticity of each pixel results from a local average of neighbouring galaxy shapes, which can have different redshifts;

(iii) the ellipticity field used by {\tt SaWLens} is a combined catalog of measurements from space- and ground-based telescopes, which probe different galaxy redshift ranges. The measurements of these two catalogs (see Ref.~\cite{Merten:2014wna}) are corrected for this, but assuming the fiducial cosmology;

(iv) in the strong lensing regime, each pixel is associated with the redshift of the multiple images contained within it, which can be different in different multiple image systems for the same cluster and also different from the galaxy populations used in the weak lensing measurements.

To get a feeling for the size of our approximation, we show in Fig.~\ref{fig:z} the $z_s$ dependence of the factor $\Upsilon$ (Eq.~(\ref{eq:factor})) for the Galileon (red) and Nonlocal (blue) models, and a $\lcdm$ model with $\Omega_{m0} = 0.4$ (black). The quantity $\Upsilon$ encapsulates all of the dependence on $z_s$ in the $\chi^2$ minimization used to estimate the cluster parameters. For illustrative purposes, we choose $z_d = 0.35$. This corresponds roughly to the mean redshift of the CLASH clusters (cf.~Table~\ref{table:d}), although the exact value is not important for the discussion here. We note that what is relevant is the slope of the curves and not their absolute value. Consider for the sake of argument an extreme case where the source galaxies are distributed between $z_s = [1,3]$, but that we choose to use $z_s = 2$ in Eq.~(\ref{eq:factor}). Focusing on the case of the Galileon model, we have that $\Upsilon(z_s =1) \approx 1.019$, $\Upsilon(z_s =2) \approx 1.011$ and $\Upsilon(z_s =3) \approx 1.007$. These values differ by no more than $\approx 1\%$, and hence our choice of $z_s$ should not lead to serious biases in the results. The error would be even smaller in the Nonlocal model or $\lcdm$ with $\Omega_{m0} = 0.40$, since in these cases the $\Upsilon(z_s)$ curves are shallower than in the Galileon case. The error of neglecting the redshift distribution becomes smaller for higher values of $z_s$, for which the curves in Fig.~\ref{fig:z} become visibly flatter. This is relevant for strongly lensed systems, which tend to be associated with galaxies at higher redshifts.

In cluster weak lensing studies, it is common to determine an effective source galaxy redshift, $z_{s,\rm eff}$, defined as 
\bq\label{eq:zeff}
\frac{D_{ds}}{D_{s}}\left(z_{s,\rm eff}\right) = \left<\frac{D_{ds}}{D_{s}}\right>,
\eq
where $\left<{D_{ds}}/{D_{s}}\right>$ is an average over all source galaxies. Reference \cite{Merten:2014wna} quotes $z_{s, \rm eff}$ values for the CLASH clusters (see also Ref.~\cite{Umetsu:2014vna}). For example, Abell 209 ($z_d = 0.206$) has $\left<{D_{ds}}/{D_{s}}\right> = 0.75 \pm 0.04$ ($1\sigma$), which corresponds to $z_{s, \rm eff} = 1.03^{+0.25}_{-0.15}$ (this estimate comes from Table 3 of Ref.~\cite{Umetsu:2014vna}). This uncertainty on $z_{s, \rm eff}$ is much smaller than our rather extreme example above ($z_s = 2\pm1$), which further convinces us that the approximation of Eq.~(\ref{eq:kapparelation}) is a good one. For completeness, we note that the determination of these values of $z_{s, \rm eff}$ involves knowledge of the background cosmology, and hence they are also model dependent. However, again taking Abell 209 as an example, in the Galileon model one has $D_{ds}/D_s(z_{s, \rm eff} = 1.03) = 0.76$, which is well within the uncertainty ($\pm 0.04$) quoted above for this cluster. We can therefore neglect this model dependency and use the values of $z_{s, \rm eff}$ listed in Ref.~\cite{Merten:2014wna}. In particular, in our $\chi^2$ minimization, we shall use the effective source redshift values found for the background sources of the ground-based ellipticity measurements, which we list in Table~\ref{table:d}.

To summarize this discussion, although Eq.~(\ref{eq:kapparelation}) is only approximate, the results shown in Fig.~\ref{fig:z} suggest that our results are insensitive to the exact choice of $z_s$.

%We shall nevertheless return to this discussion in Sec.~\ref{sec:zsimpact}, where we reassess the impact of $z_s$ on our results.

%===========================================================================================================================================%
% CLASH subtleties
%===========================================================================================================================================%

\subsection{Other subtleties in using cluster lensing data to test gravity}

Before proceeding further into estimating the CLASH cluster masses in modified gravity models, we discuss some other subtle issues that may arise when combining current lensing modelling techniques with modified gravity. Although it turns out these other issues do not play a direct role in the results of this paper, we believe such a discussion is instructive and leads to a clearer and broader understanding of the results of this and other work in the literature.

\subsubsection{Parametric vs. nonparametric analysis}\label{sec:issues}

The non-parametric reconstruction of the lensing potential used in this paper builds solely upon the observed lensing constraints, without making any assumptions about the mass distribution of the cluster. Such a model-independent \footnote{Apart from the issue of the fiducial cosmological background model discussed above.} method is particularly well suited to modified gravity studies. Consider the alternative scenario of a parametric approach. In this case one starts by making an {\it Ansatz} about the mass distribution in the cluster. Typically, this can involve describing the main dark matter distribution using a single (or more in the case of mergers) NFW profile. Then, one could also model substructure by identifying the position of the most massive cluster galaxies and assigning them a given density profile. (see e.g.~Refs.~\cite{Kneib:1995hh, Broadhurst:2004hu, Smith:2004wr, Halkola:2006pq, Jullo:2007up, 2009MNRAS.396.1985Z, 2010PASJ...62.1017O, 2013ApJ...765...24N, Jullo:2013fty, Johnson:2014swa, Monna:2013eia, 2011ApJ...738...41U, 2012MNRAS.420.3213O, Jauzac:2014nga, Jauzac:2014jha, Jauzac:2014xwa}). The free parameters of such a mass model  are then iterated over until the lensing constraints are satisfied. In the context of modified gravity there are at least two subtle issues associated with such a parametric lensing analysis. First, in order to compute the lensing effects due to the postulated mass distribution one must assume a theory of gravity: for the same mass distribution, different models of gravity could induce different lensing effects. Parametric methods are therefore biased towards the assumed theory of gravity. Second, the lensing properties of a given point in the cluster field are determined by the sum of the lensing signal predicted by each element of the mass model (main halo plus the substructures). This superposition is valid in GR (which is linear in the Newtonian limit), but not necessarily in alternative (typically nonlinear) models of gravity. These issues can be circumvented if one reconstructs directly the lensing potential and its derivatives but not the mass distribution. It is for this reason that we choose to use the {\tt SaWLens} results of Ref.~\cite{Merten:2014wna} in our analysis.

For clarity, it is worth pointing out that the problem with parametric mass modelling is less severe if it is only applied to the strong lensing part of the data analysis. In this regime (well within $R_{200}$), the effects of the modifications to gravity in a model like the Galileon may be small by virtue of the screening mechanism. As a result, the assumptions made in the data analysis may turn out to be a good approximation. This is the case for some of the Galileon-type models explored in Refs.~\cite{Narikawa:2013pjr, 2012JCAP...05..016N}, which were compared to (parametric) cluster data from Refs.~\cite{2011ApJ...738...41U, 2012MNRAS.420.3213O}. However, it is reasonable to expect that the screening efficiency is different in different models of gravity. For instance, the Nonlocal gravity model predicts a constant enhancement of the gravitational strength on all relevant scales. One must therefore be always cautious and check for the compatibility of the data analysis with the specific theory of gravity to be tested.

\subsubsection{Interpretation of stacked cluster lensing profiles}
To overcome systematic effects due to intervening structure, cluster substructure and cluster asphericity, it has become common to build average (stacked) lensing profiles by using cluster lensing data from independent lines of sight \cite{2011ApJ...729..127U, 2011ApJ...738...41U, 2012MNRAS.420.3213O}. The averaged profiles are then fitted again to infer an average mass and concentration that characterizes the stack. From a conceptual point of view, the same procedure can be applied assuming modified gravity models. Here, we comment that the interpretation of the stacked data may be somewhat more complex due to the effects of modified gravity. Consider for simplicity the stacking of the convergence radial profiles of $N$ clusters at redshifts $z_{1..N}$ with mass and concentration values $M_{1..N}$ and $c_{1..N}$, respectively. The background galaxies can be assumed to lie at the same source redshift. For instance, Ref.~\cite{2011ApJ...738...41U} stacks four massive clusters by co-adding (with some weighting) their profiles. The resulting mean profile is then refitted to determine a mean mass and concentration of the stack. Now consider fitting such a stack to two gravity models which display different time evolution for an unscreened gravitational strength. For these two models, clusters located at different redshifts would contribute differently to the mean mass/concentration estimate since their lensing signal is amplified differently. For such a scenario, an interesting analysis would be to split the stack into smaller ones binned by cluster redshift and check for differences in the resulting mean mass/concentration of the smaller stacks. The situation becomes even more complex (but interesting) in models with screening, due to its scale-dependence, whose efficiency is in general redshift dependent as well.

We stress that the above issues do not pose a serious problem to using stacked data to test modified gravity, but simply that the extra physics can enrich the interpretation of the results. In this paper, however, we shall not be concerned with these issues since we fit each of the CLASH clusters individually.

\section{Results: lensing mass estimates}\label{sec:results}

In this section, we present our main results for the mass and concentration estimates of the CLASH clusters in the three models of gravity we consider. We also discuss the impact of our results in the context of recently proposed observational methods to test gravity on large scales.

%===========================================================================================================================================%
% RESULTS OMEGA
%===========================================================================================================================================%

\subsection{The impact of $\Omega_{m0}$ on cluster lensing mass estimates}

\begin{figure*}
	\centering
	\includegraphics[scale=0.39]{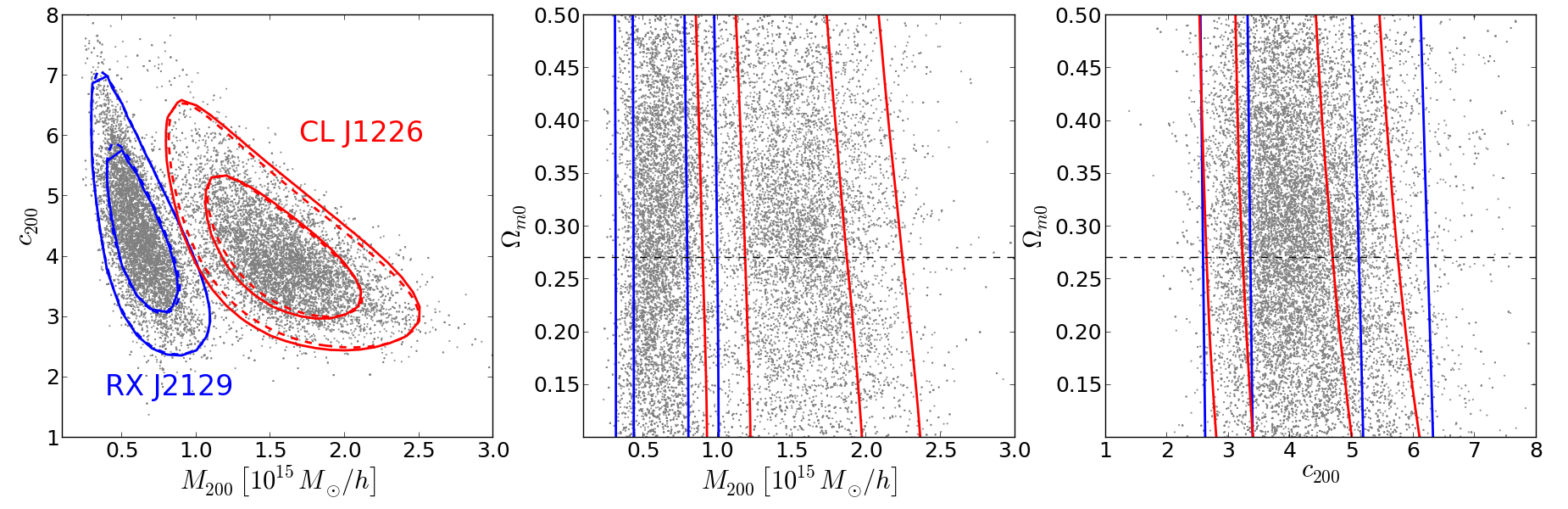}
	\caption{Two-dimensional marginalized constraints on the $c_{200}-M_{200}$ (left), $\Omega_{m0}-M_{200}$ (middle) and $\Omega_{m0}-c_{200}$ (right) planes, for the CLASH clusters RX J2129 (blue) and CL J1226 (red), assuming a $\lcdm$ cosmology. The solid contours depict the marginalized $68 \%$ and $95\%$ confidence limits when all three parameters are varied. The dashed contours in the left panel are the same as the solid ones, but with the cosmological matter density held fixed at $\Omega_{m0} = 0.27$ (barely noticeable in the case of RX J2129). The gray dots show the points accepted in the Monte Carlo chains built by the {\tt CosmoSIS} code.}
\label{fig:varyOm0}\end{figure*}

\begin{figure}
	\centering
	\includegraphics[scale=0.345]{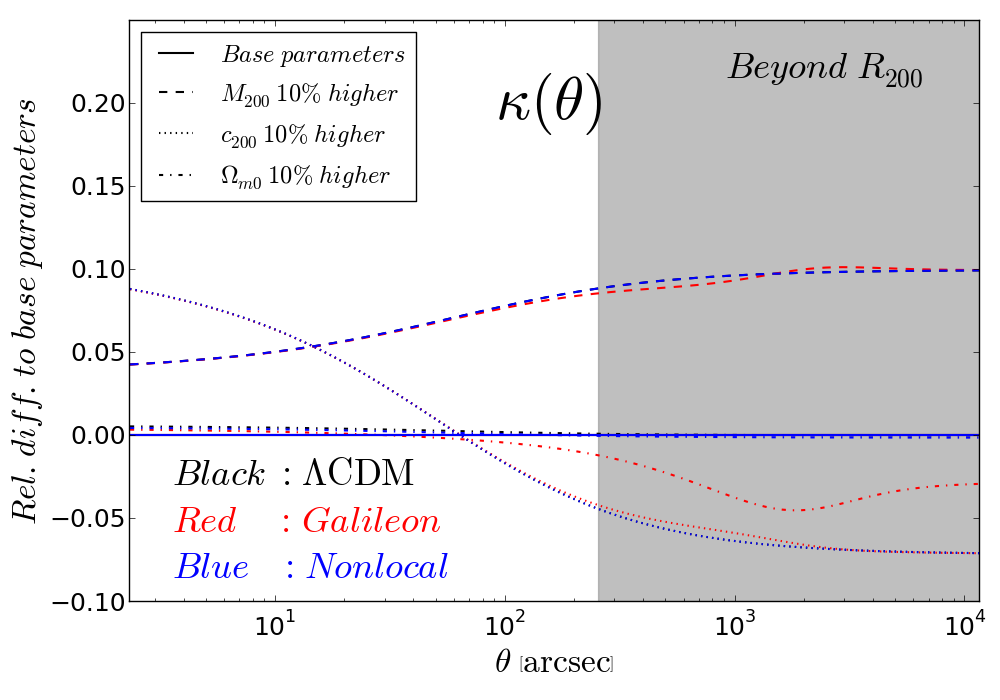}
	\caption{Lensing convergence $\kappa(\theta)$ for $\lcdm$ (black), Galileon gravity (red) and Nonlocal gravity (blue) models, plotted as the difference relative to their respective models with base parameters $M_{200} = 5 \times 10^{14} M_{\odot}/h$, $c_{200} = 4$ and $\Omega_{m0} = 0.27$. The blue and black curves are almost overlapping. We assume $z_d = 0.5$ and $z_s = 1.0$. The dashed, dotted and dot-dashed lines correspond, respectively, to cases for which the values of $M_{200}$, $c_{200}$ and $\Omega_{m0}$ are enhanced by $10\%$. To guide the eye, the shaded region indicates the radial scales beyond $R_{200}$ in the base $\lcdm$ model.}
\label{fig:varyparams}\end{figure}

Figure \ref{fig:varyOm0} shows the constraints in the $c_{200}$-$M_{200}$ (left), $\Omega_{m0}$-$M_{200}$ (middle) and $\Omega_{m0}$-$c_{200}$ (right) planes for the RX J2129 (blue) and CL J1226 (red) CLASH clusters, assuming a $\lcdm$ model. The solid contours show the parameter constraints obtained using a three-dimensional grid search. The gray dots correspond to the points accepted in the chains of a Monte Carlo exploration of the parameter space using the recently developed and publicly-available {\tt CosmoSIS}\footnote{The bitbucket webpage of the {\tt CosmoSIS} code is https://bitbucket.org/joezuntz/cosmosis/wiki/Home .} package \cite{Zuntz:2014csq}. {\tt CosmoSIS} is a highly modular parameter estimation code, to which we have added likelihood modules based on the convergence profiles of each of the 19 CLASH clusters. Given that our grid code and the MCMC sampler in {\tt CosmoSIS} are independent ways of sampling the parameter space, it is reassuring that their results agree as shown in Fig.~\ref{fig:varyOm0}. We will be making the CLASH likelihood modules developed in this paper publicly available in future releases of the {\tt CosmoSIS} standard library.

The middle and right panels of Fig.~\ref{fig:varyOm0} show that the data from each cluster does not place any meaningful constraints on $\Omega_{m0}$ (this conclusion holds for all of the other CLASH clusters as well). Furthermore, there is no clear degeneracy between $\Omega_{m0}$ and the cluster parameters $M_{200}$, $c_{200}$, which indicates that the constraints on the mass and concentration are unbiased w.r.t.~a particular choice of $\Omega_{m0}$. This point is also illustrated by the dashed contours in the left panel of Fig.~\ref{fig:varyOm0}, which show the constraints when the cosmological matter density is held fixed at $\Omega_{m0} = 0.27$. The comparison of the solid and dashed contours shows that there is no significant deterioration of the constraints when one allows $\Omega_{m0}$ to be a free parameter. The cases of CL J1226 (red) and MACS J0744 (not shown) are those  for which the deterioration is the most pronounced (although still very small). For the rest of the clusters, the two sets of contours (solid and dash) are barely distinguishable, just as in the case of RX J2129 (blue). This illustrates that, {\it in observational determinations of cluster masses from lensing, the impact of assuming a specific value of $\Omega_{m0}$ (as is often done in the literature) is negligible} \footnote{The changes in the constraints when one allows $\Omega_{m0}$ to vary are so small that they are of the same magnitude as the error of the approximation of Eq.~(\ref{eq:kapparelation}), which is expected to be negligible in any case, as discussed in Sec.~\ref{sec:validity}.}.

To help understand the above result (and others that will follow), it is instructive to look at the impact of $M_{200}$, $c_{200}$ and $\Omega_{m0}$ on the lensing convergence profiles. This is shown in Fig.~\ref{fig:varyparams} for $\lcdm$ (black), Galileon gravity (red) and Nonlocal gravity (blue). For all models:

\setlength{\leftmargini}{10pt}\begin{itemize}

\item an increase in $c_{200}$ (keeping all other parameters fixed) tilts the convergence profile, boosting its amplitude in the inner regions and lowering it in the outer regions of the halo; 

\item larger mass values enhance the lensing convergence on all scales shown, with the effect being slightly more pronounced at large radii\footnote{{Note that for $\theta \gtrsim 10^2 {\rm arcsec}$, the effects of increasing halo mass start to become degenerate with the effects of decreasing the concentration. Strong lensing analysis, which probes the inner most regions of the cluster, therefore helps to break this {\it concentration-mass degeneracy}. Reference \cite{Umetsu:2014vna} also determines the lensing profiles of some of the CLASH clusters used in Ref.~\cite{Merten:2014wna}. However, the latter includes strong-lensing constraints in the analysis, which allows for more accurate concentration estimations. This is why we choose to work with the results of Ref.~\cite{Merten:2014wna}.}};

\item A boost in $\Omega_{m0}$ has three main effects. First, it increases the value of $\rho_c(z)$ at the cluster redshift, which effectively increases the density of the halo for fixed $M_{200}$ (recall the discussion of Sec.~\ref{sec:fixedparams}). This should boost the convergence at small radii. Second, an increase in $\Omega_{m0}$ also has an impact on the cosmological distances that enter $\Sigma_c$, which has the overall effect of lowering the amplitude of $\kappa$, by a constant factor. Third, a change in $\Omega_{m0}$ also changes the angular diameter distance to the cluster, $D_d$, which causes horizontal shifts in the lensing convergence, when plotted as a function of the observed angular scales $\theta = D_d/r$. 

\end{itemize}

The net effect of varying $\Omega_{m0}$ results in almost no visible shift (dot-dashed) in the amplitude of $\kappa$ for the radial scales that are better probed by the CLASH radially binned profiles, $\theta \lesssim 500-700\ {\rm arcsec}$ (cf.~Fig.~\ref{fig:kappas}). This is why the constraints on $\Omega_{m0}$ in Fig.~\ref{fig:varyOm0} are so weak. We note also that since the effect of $\Omega_{m0}$ is always subdominant w.r.t.~the effects of varying $M_{200}$ and $c_{200}$ (in terms of fractional change), it is unlikely that future lensing data from experiments such as EUCLID \cite{2011arXiv1110.3193L} and LSST \cite{2012arXiv1211.0310L} will change this conclusion.

From hereonin, we fix the cosmological matter density to be $\Omega_{m0} = 0.27$ for all our models. In this way, the $\lcdm$ model becomes the fiducial one used in Ref.~\cite{Merten:2014wna}. This value is also consistent with the CMB observational constraints for the Cubic Galileon and Nonlocal models as found, respectively, by Ref.~\cite{Barreira:2014jha} and Ref.~\cite{Dirian:2014bma}.

%===========================================================================================================================================%
% RESULTS MODIFIED GRAVITY
%===========================================================================================================================================%

\subsection{Cluster lensing masses in Galileon and Nonlocal gravity}\label{sec:galimasses}

\begin{figure*}
	\centering
	\includegraphics[scale=0.39]{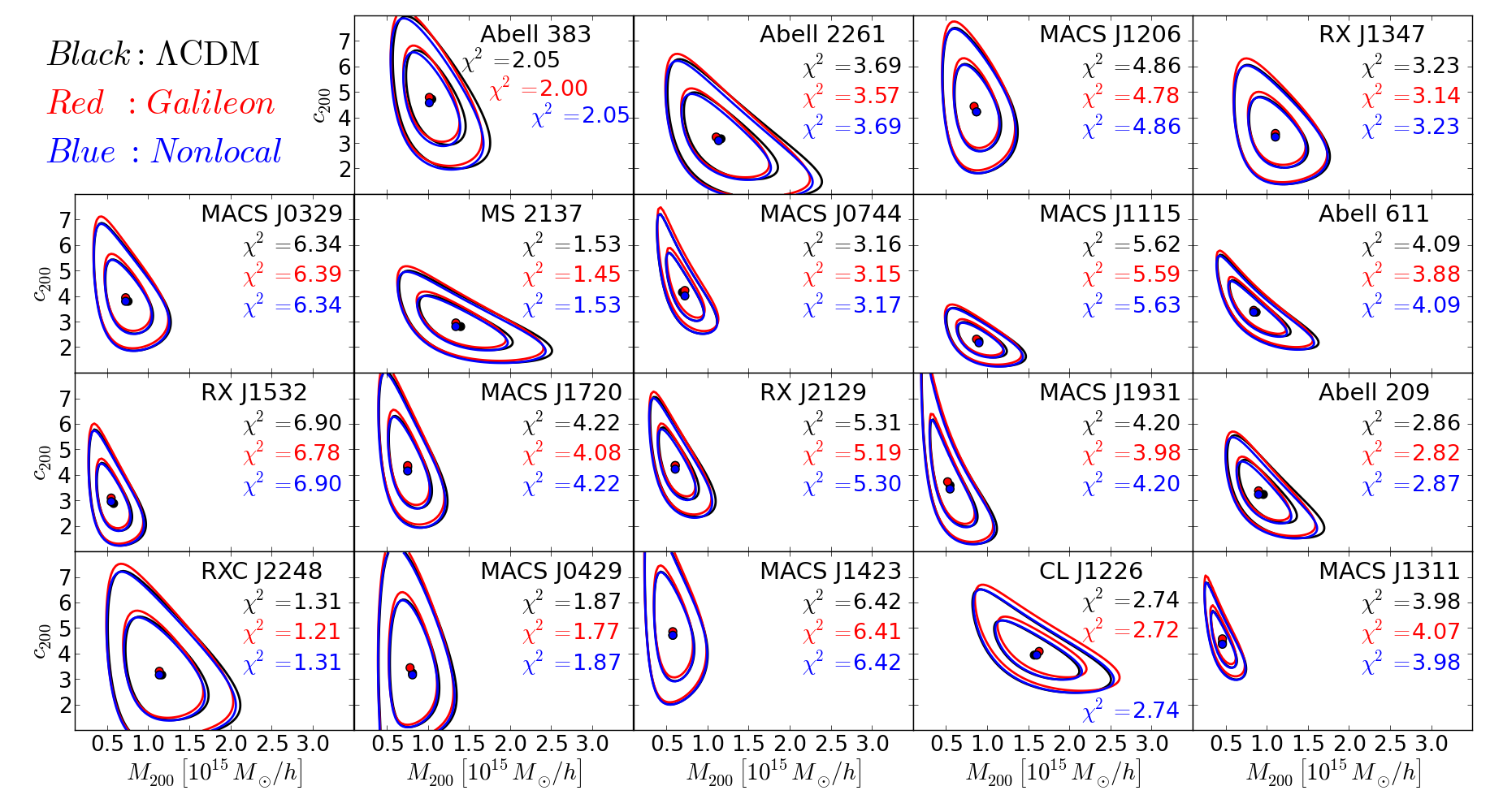}
	\caption{Two-dimensional $68\%$ and $95\%$ confidence limits on the $c_{200}-M_{200}$ plane for all of the CLASH clusters assuming $\lcdm$ (black), Galileon gravity (red) and Nonlocal gravity (blue). The position of the best-fitting points is marked by the dots, and their respective $\chi^2$ values are shown in each panel.}
\label{fig:contours}\end{figure*}

\begin{figure*}
	\centering
	\includegraphics[scale=0.39]{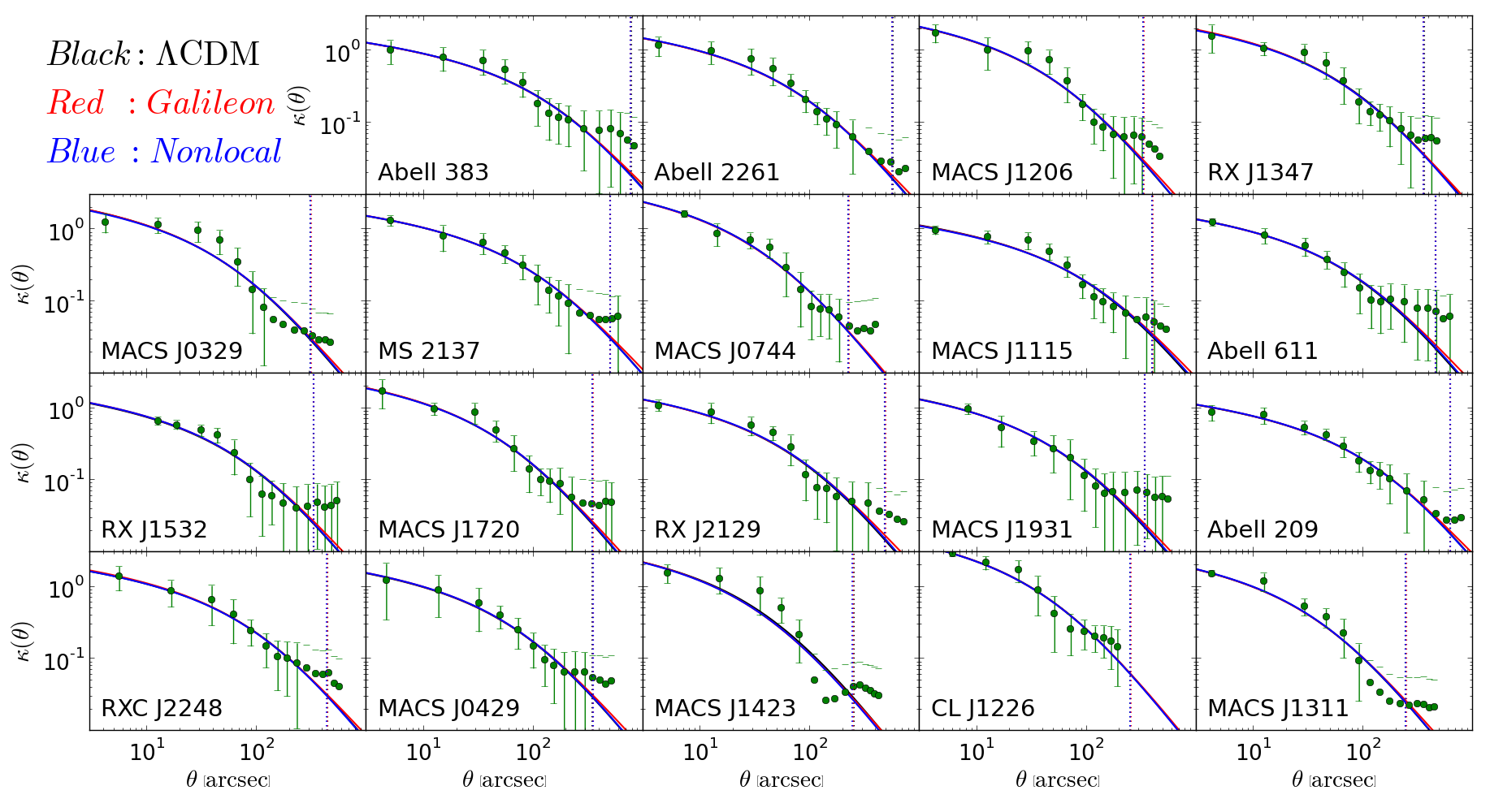}
	\caption{Best-fitting lensing convergence profiles, $\kappa(\theta) = \Upsilon\kappa_{\infty}$, obtained for all of the CLASH clusters assuming $\lcdm$ (black), Galileon gravity (red) and Nonlocal gravity (blue) . The green dots are the radially binned data as described in Ref.~\cite{Merten:2014wna} and the errorbars are the square root of the diagonal entries of the covariance matrix of the data. To guide the eye, the dotted vertical lines indicate the inferred values of $R_{200}$, which are barely distinguishable for the three models.}
\label{fig:kappas}\end{figure*}

\begin{figure}
	\centering
	\includegraphics[scale=0.345]{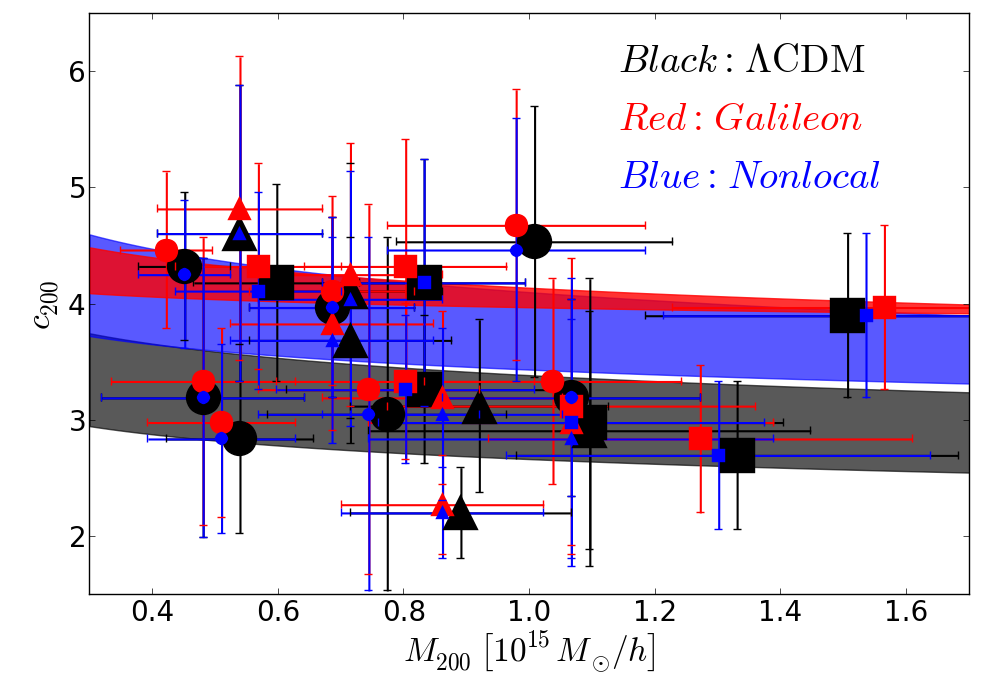}
	\caption{Concentration-mass relation of the CLASH clusters assuming $\lcdm$ (black), Galileon gravity (red) and Nonlocal gravity (blue). The errorbars indicate the marginalized $68\%$ confidence limits. We use different symbols for different clusters to facilitate the identification of which cluster is which across the three models. The shaded bands indicate the mean concentration-mass relations from N-body simulations between $z = 0.66$ (lower bound) and $z=0$ (upper bound) found for $\lcdm$ (gray) in Ref.~\cite{Duffy:2008pz}, Cubic Galileon model (red) in Ref.~\cite{Barreira:2014zza} and Nonlocal model (blue) in Ref.~\cite{Barreira:2014kra}.}
\label{fig:cM}\end{figure}

Figure \ref{fig:contours} shows the constraints on the $c_{200}$-$M_{200}$ plane obtained for each of the CLASH clusters in $\lcdm$ (black), Galileon (red) and Nonlocal gravity (blue) cosmologies. The dots indicate the position of the best-fitting values. The best-fitting lensing convergence profiles are shown in Fig.~\ref{fig:kappas} (what is shown is $\Upsilon\kappa_{\infty}(\theta)$). The concentration-mass relation of the CLASH clusters for the three models is shown in Fig.~\ref{fig:cM}, together with results from N-body simulations \cite{Duffy:2008pz, Barreira:2014zza, Barreira:2014kra}. First, we note that our cluster mass and concentration estimates for $\lcdm$ are in agreement with those obtained in Ref.~\cite{Merten:2014wna}. Second, these three figures all show that the constraints on the cluster parameters are, within errorbars, the same in the three cosmological models. {Although there are tiny differences in the resulting best-fitting values of $M_{200}$ and $c_{200}$ for the three models ($\lesssim 5\%$), they all lie well within the $1\sigma$ limits (whose precision varies within $\sim 50\%-80\%$)}. The shapes of the contours are also remarkably similar and the goodness-of-fit is essentially the same in all models, as can be seen by comparing the respective $\chi^2$ values in Fig.~\ref{fig:contours}. In Fig.~\ref{fig:kappas}, one notes that for almost all of the clusters, the best-fitting convergence profiles underpredict the data points at large angular scales (although well within the errorbars). However, close to the edge of the clusters, the contribution from the surrounding large scale structure may have a non-negligible impact. This can partly explain why the data points tend to go up at large scales, as investigated, for instance, in Ref.~\cite{2011MNRAS.414.1851O}.

The shaded bands in Fig.~\ref{fig:cM} show the best-fitting mean concentration-mass relations found in N-body simulations for the $\lcdm$ (gray) model in Ref.~\cite{Duffy:2008pz} \footnote{See Fig.~9 of Ref.~\cite{Merten:2014wna} for the comparison of the CLASH $c_{200}$-$M_{200}$ relation in $\lcdm$ with other relations in the literature.}, the Cubic Galileon model (red) in Ref.~\cite{Barreira:2014zza} and the Nonlocal model (blue) in Ref.~\cite{Barreira:2014kra}. In these bands, the lower and upper bounds correspond, respectively, to the relations at $z = 0.666$ ($a = 0.60$) and $z = 0$ ($a = 1$) (this redshift range is approximately that of the CLASH clusters). Figure \ref{fig:cM} shows that there is good agreement between the simulation results and the concentration/mass estimates of the CLASH clusters in the three models of gravity. However, there are a number of issues that prevent a direct comparison between the simulation results and the estimated concentration and mass values. First, the shaded bands of the Galileon and Nonlocal models in Fig.~\ref{fig:cM} have been extrapolated to masses larger than the mass range used to fit the best-fitting concentration-mass relations in the simulations of Refs.~\cite{Barreira:2014zza} and \cite{Barreira:2014kra}. Second, the concentration-mass relation was fitted using all haloes, without applying any selection criteria to consider only relaxed ones \cite{Neto:2007vq}. This may be particularly relevant for the CLASH clusters, which are characterized by regular X-ray surface brightness morphologies \cite{2012ApJS..199...25P}, and are therefore expected to be relaxed and close to virial equilibrium (see also Refs.~\cite{Schaller:2014uwa, Schaller:2014gwa} for a recent discussion on the impact of baryonic processes in the density profiles of clusters). Third, the concentration-mass relation in the simulations was obtained by fitting NFW profiles to the three-dimensional spherically averaged mass distribution of the haloes, whereas the symbols in Fig.~\ref{fig:cM} are the values obtained by also assuming spherical symmetry, but fitting to two-dimensional (projected) lensing convergence profiles (see e.g.~Sec.6.2 of Ref.~\cite{Merten:2014wna} for an analysis of the impact of this {\it projection bias} in the CLASH sample). {Finally, the upper and lower bounds of the bands correspond to the mean relation found in the simulations, but the intrinsic scatter around the mean concentration-mass relation should also be taken into account.} Nevertheless, to guide the eye, we opted to keep the simulation results in Fig.~\ref{fig:cM}, but advise that further work is needed before performing a more thorough comparison {(see the analysis of Ref.~\cite{Meneghetti:2014xna} in $\Lambda$CDM models for an illustration of the steps to follow).}

% The concentration-mass relation of the Galileon and Nonlocal models lie above that of $\lcdm$, and inspection hints also that, on average, the bands of these models slightly overpredict the symbols.

The left panel of Fig.~\ref{fig:focus} shows the best-fitting lensing convergence for all of the CLASH clusters in the Galileon (red) and Nonlocal (blue) cosmologies, plotted as the respective difference to the best-fitting profiles in $\lcdm$. As expected from the above results, on the scales that are probed by the CLASH data, $\theta \lesssim 500-700\ {\rm arcsec}$, the three models are in very good agreement. In the case of the Galileon model, this is because the screening is very effective on these scales inside $R_{200}$ (Ref.~\cite{Falck:2015rsa} finds a similar screening efficiency inside $R_{200}$ for DGP gravity, which employs also the Vainshtein mechanism). This can be noted by comparing the enhancement in the amplitude of $\kappa$ on larger scales, where the screening becomes less efficient. In the case of the Nonlocal model, although the modifications to the gravitational strength are not screened, they are not strong enough to have a significant impact on the lensing convergence profiles. {\it We therefore conclude that, for the case of the CLASH clusters analysed here, the impact of modifying the lensing gravitational potential according to Cubic Galileon or Nonlocal gravity  is completely negligible in the estimation of their lensing masses.} 

Before this paper, there have been other works investigating the impact of Galileon-like effects on cluster lensing profiles. Reference~\cite{2012JCAP...05..016N} used a parametrization of the fifth force, which was constrained using the stacked cluster lensing shear profiles from Refs.~\cite{2011ApJ...738...41U, 2012MNRAS.420.3213O}. Their parametrization encompasses the Galileon model studied here, and for this case, there is good agreement with our conclusions. Reference \cite{Narikawa:2013pjr} performed similar investigations, but took as a test case a model inspired by massive gravity. The authors do not consider the time evolution of the cosmological background and their equations of motion include higher order terms than the Galileon model studied here, which prevents a direct comparison with our results. A particularly interesting feature described in Ref.~\cite{Narikawa:2013pjr} is that some model parameters predict a "dip" in the amplitude of the convergence profiles, which happens for $r \lesssim R_{200}$ (see e.g.~Fig.~4 of Ref.~\cite{Narikawa:2013pjr}). These scales are sufficiently well probed by the lensing data, which allows some of these specific models to be ruled out already. Such features, however, do not show up in our convergence profiles for the Cubic Galileon model (cf.~Fig.~\ref{fig:kappas}), {but which is a different model from the one studied in Ref.~\cite{Narikawa:2013pjr}}. In the context of $f(R)$ models, although the lensing signal is not modified directly, it can be via modified mass distributions. For instance, the enhanced dynamical potential in these models boosts the accretion rate of matter onto the clusters, which results in an excess of mass in their infall region. This can be probed with lensing measurements, as was done in Ref.~\cite{2012PhRvD..85j2001L}. Since the Galileon and Nonlocal models also directly modify the dynamical potential, then in principle, similar investigations can be performed. In the case of these models, the lensing signal should be amplified both by the excess of mass that surrounds the cluster, and by the intrinsically enhanced lensing effects. A detailed investigation of this is, however, beyond the scope of the present paper.

%===========================================================================================================================================%
% CONNECTION TO TESTS
%===========================================================================================================================================%

\begin{figure*}
	\centering
	\includegraphics[scale=0.39]{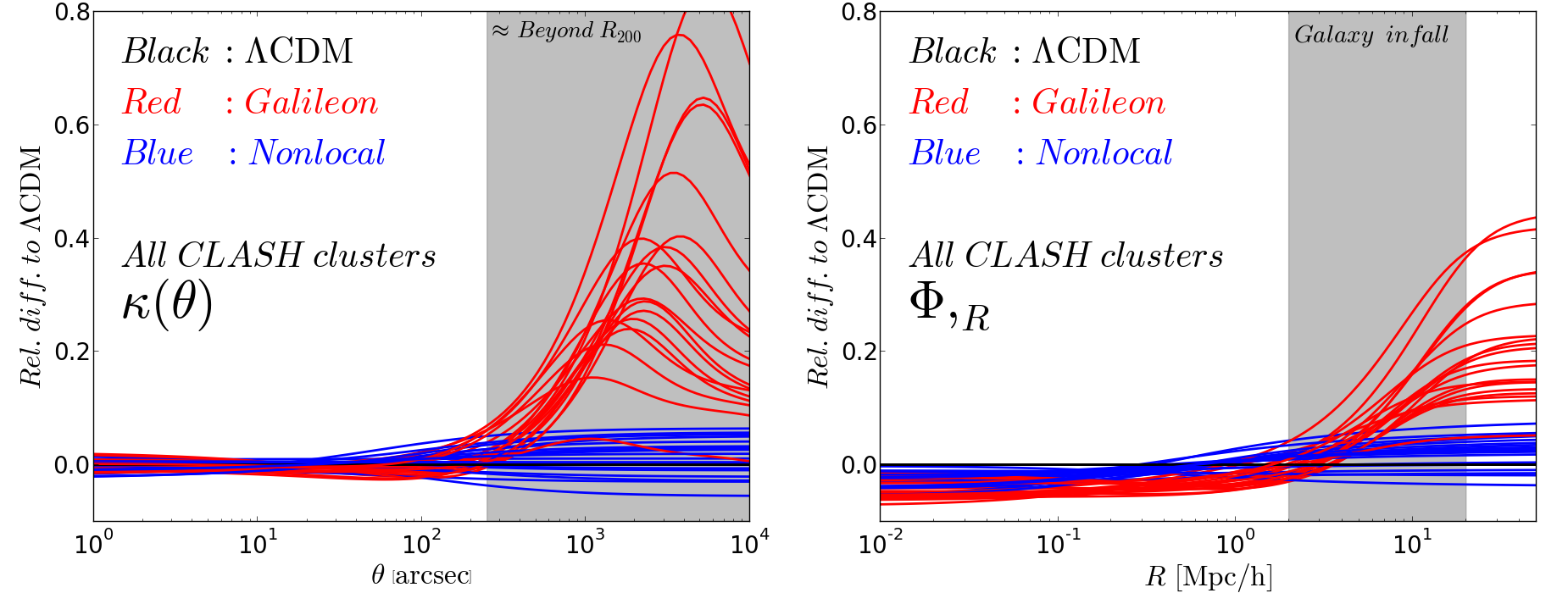}
	\caption{(Left) Best-fitting lensing convergence profiles, $\kappa(\theta) = \Upsilon\kappa_{\infty}$, for the all of the CLASH clusters in the Galileon (red) and Nonlocal (blue) gravity models, plotted as the difference relative to the best-fitting profiles in $\lcdm$ (black). To guide the eye, the shaded band represents approximately the regions that lie beyond $R_{200}$ for all clusters. (Right) Same as the left panel, but for the total force profile $\Phi,_R$. The shaded band  encloses the scales $2{\rm Mpc}/h \lesssim R \lesssim 20{\rm Mpc}/h$ which are approximately those associated with the infall of surrounding galaxies.}
\label{fig:focus}\end{figure*}

\subsection{The connection with tests of gravity}\label{sec:test}

\subsubsection{Dynamical masses from the phase-space density around massive clusters}

Recently, the authors of Refs.~\cite{2012PhRvL.109e1301L, Lam:2013kma, Zu:2013joa, Wilcox:2015kna} have proposed methods to test the law of gravity on ${\rm Mpc}$ scales by using information from the galaxy velocity field in the infall regions around massive clusters (see also Ref.~\cite{Hellwing:2014nma}). These techniques were designed with models of gravity that modify the dynamical potential (i.e.~that felt by nonrelativistic objects like galaxies), but do not modify the lensing potential (i.e.~that felt by relativistic particles like photons). Popular models such as $f(R)$ and DGP gravity fall in the above category, and as such, the lensing mass estimates, $M_{\rm len}$, for these models would automatically be the same as in GR. On the other hand, the velocity dispersion of surrounding galaxies as they fall towards the clusters would be affected by the modifications to gravity. Therefore, if one would interpret these observations assuming GR, then one would infer dynamical masses, $M_{\rm dyn}$, which are different from those estimated using lensing. A mismatch in the estimates of the lensing and dynamical masses would therefore be a smoking gun for modified gravity \cite{2009arXiv0907.4829S, 2010PhRvD..81j3002S, 2011PhRvL.107g1303Z} (see, however, Ref.~\cite{Hearin:2015lca} for a discussion of how complications associated with assembly bias could affect these tests).

The merit of the test of gravity described above becomes less clear when applied to models that also modify the lensing potential. Consider, for simplicity, a model that boosts the dynamical and lensing potential by the same constant factor, $\alpha>1$, i.e.~ $\Phi_{\rm dyn} = \Phi_{\rm len} \sim \alpha \Phi^{\rm GR}$. In such a model, the mass of a cluster inferred from the surrounding galaxy velocity field would be biased low w.r.t.~GR. This is because, due to the enhanced gravitational strength felt by the galaxies, the cluster does not need to be as massive as in GR to accelerate the galaxies by the same amount. Following the same reasoning, the lensing mass estimates would also tend to be biased low compared to GR: due to the fifth force felt by the photons, the cluster can be less massive to induce lensing effects of the same magnitude. In such a model, both $M_{\rm dyn}$ and $M_{\rm len}$ shift in the same direction. This therefore makes it harder to tell the two values apart and hence, harder to detect a signature of modified gravity.

{The Galileon model also modifies the lensing potential, but adds complexity to the case described above in the sense that the modifications to gravity are scale dependent with screening inside the cluster radius. Just outside the cluster radius, the screening becomes less efficient and the fifth force significantly boosts the lensing convergence, as shown in Fig. 9. Although these larger scales are not accurately probed by the current cluster lensing data, they correspond roughly to the regions associated with galaxy infall, $2 {\rm Mpc}/h \lesssim r \lesssim 20 {\rm Mpc}/h$. For these radial scales, the right panel of Fig. 9 shows that the total force profile which surrounds the CLASH clusters in a Galileon cosmology can be up to $10\%-40\%$ higher than in $\lcdm$. As a result, galaxies located at these distances from the cluster center should feel the boost in the total force, which should translate into their velocity distribution. On these scales, both the lensing and dynamical masses would be different in a Galileon cosmology compared to $\lcdm$. Inside the cluster radius, on the scales that are probed by the CLASH data, $\theta \lesssim 500-700\ {\rm arcsec}$, the left panel of Fig. 9 shows that the differences in the convergence profiles compared to those in $\lcdm$ are small enough for the mass estimates to be almost the same in the two models. Therefore, inside the cluster radius, this leaves us with a similar picture to that in $f(R)$ or DGP models: the lensing mass estimates are not affected by the modifications to gravity, but dynamical mass estimates using infalling galaxies are changed. We therefore conclude that, {\it despite it being a model that modifies the lensing potential, the fact that dynamical and lensing mass estimates are sensitive to radial scales of different screening efficiency allows the Cubic Galileon model to be tested by the methods proposed in Refs.}~\cite{2012PhRvL.109e1301L, Lam:2013kma, Zu:2013joa}.}

In the case of the Nonlocal model, although the lensing mass estimates are also practically the same as in $\lcdm$, the enhancement of the force profile on scales $2 {\rm Mpc}/h \lesssim r \lesssim 20 {\rm Mpc}/h$ is kept below the $\sim 5\%$ level. This makes it more challenging for this model to be tested by these methods.

\subsubsection{Galaxy-galaxy lensing}

The left panel of Fig.~\ref{fig:focus} also shows that although the convergence profiles are very close in the three models for $R \lesssim R_{200}$, they can be visibly higher (by $\sim 20-80\%$) in the Galileon model on larger scales. The enhanced lensing signal outside dark matter haloes in Galileon-like models has been analysed by Refs.~\cite{2011PhRvL.106t1102W, Park:2014aga}, but in the context of theories that emerge from massive gravity scenarios \cite{Gabadadze:2009ja, deRham:2009rm, deRham:2010kj, deRham:2010gu, deRham:2010ik, deRham:2010tw}. In particular, the authors investigate the possibility of such a signal being detected in galaxy-galaxy lensing observations (see e.g.~Ref.~\cite{2013MNRAS.432.1544M}). The latter can be measured by cross-correlating the position of foreground galaxies (the lenses) with their background shear field. Our results in Fig.~\ref{fig:focus} are in good qualitative agreement with the solutions explored in Refs.~\cite{2011PhRvL.106t1102W, Park:2014aga}. For instance, we also find the appearence of a bump in the relative difference to $\lcdm$, which we checked occurs at $\sim 10 R_{200}$. Quantitatively, the comparisons become less straightforward. On the one hand, in this paper we show the results for cluster mass scales between $\approx \left[0.5, 1.5\right] \times 10^{15} M_{\odot}/h$, which are higher than the galaxy group mass scales ($10^{13} - 10^{14} M_{\odot}/h$) probed in Refs.~\cite{2011PhRvL.106t1102W, Park:2014aga}. Moreover, our models also differ at the level of the cosmological background, exact screening efficiency and time evolution of the linearized effective gravitational strength. Nevertheless, it seems reasonable to expect that the predictions of the Galileon model studied here are also likely to be scrutinized by galaxy-galaxy lensing observations. A more detailed investigation of the model predictions for galaxy-galaxy lensing is beyond the scope of the present paper.

In the case of Nonlocal gravity, the modifications to the lensing convergence are small ($\lesssim 5\%$) on all scales, which makes it much harder to distinguish from standard $\lcdm$ with galaxy-galaxy lensing data.

\subsubsection{Weak lensing on larger scales}

The picture depicted in the left panel of Fig.\ref{fig:focus} that the lensing signal gets significantly enhanced on larger scales in the Galileon model should, in principle, also have an impact on the lensing of CMB photons. Indeed, Refs.~\cite{2012PhRvD..86l4016B, Barreira:2014jha} have shown that the amplitude of the CMB lensing potential angular power spectrum, $C_l^{\psi\psi}$, is very sensitive to the modifications to gravity in the Galileon model. To the best of our knowledge, the effect of the Nonlocal model on the CMB lensing potential power spectrum has never been investigated in detail. However, since the modifications to gravity on large scales are not as strong as in the Galileon model (cf.~middle panel of Fig.~\ref{fig:bg}), the effects on the amplitude of $C_l^{\psi\psi}$ should be less pronounced.

By the same reasoning the weak lensing cosmic shear power spectrum should also be sensitive to the modifications to gravity in the Galileon model, but less so in the Nonlocal case. Again to the best of our knowledge, cosmic shear data, such as that gathered by the CFHTLens Survey \cite{2012MNRAS.427..146H}, have never been used in direct tests of the models studied here, although Refs.~\cite{Battye:2014xna, Leonard:2015hha, Ade:2015rim} have used these data to constrain general parametrizations of modified gravity.

%===========================================================================================================================================%
% SUMMARY
%===========================================================================================================================================%

\section{Summary \& Outlook}\label{sec:summary}

We have estimated cluster lensing masses in alternative theories of gravity that modify the lensing gravitational potential. For this, we varied the mass ($M_{200}$) and concentration ($c_{200}$) of NFW haloes to fit the predicted lensing signal in modified gravity to the radially binned lensing convergence profiles obtained from non-parametric reconstructions of the lensing potential for 19 X-ray selected clusters from the CLASH survey \cite{2012ApJS..199...25P, Merten:2014wna, Umetsu:2014vna}. The methodology we adopted is similar to that first employed in Ref.~\cite{Merten:2014wna} in the context of GR.

We focused on the Cubic Galileon and Nonlocal models, which modify the gravitational law in qualitatively different ways. In the case of the Nonlocal model, the modifications to gravity can be parametrized by an effective time varying gravitational strength, which is independent of length scale (cf.~Eq.~(\ref{eq:geff}) and Fig.~\ref{fig:bg}). In the Galileon model, the gravitational law can also be parametrized by a scale-independent gravitational strength on large scales (cf.~Eq.~({\ref{eq:geffgali}) and Fig.~\ref{fig:bg}). However, close to massive bodies, the Vainshtein mechanism (manifest in nonlinearities in the equations of the model) introduces a scale dependency to the gravitational strength, which acts to suppress the amplitude of the modifications w.r.t.~standard GR/$\lcdm$. The cosmological background in both models is also modified relative to $\lcdm$ (cf.~Fig.~\ref{fig:bg}), which has an impact on the conversion between angular and radial scales, $\rho_{c}(z)$ and also on the values of $\Sigma_c$.

We paid particular attention to the compatibility of the data analysis with the modified gravity models we wished to test. Namely, we pointed out that the CLASH cluster convergence profiles obtained by Ref.~\cite{Merten:2014wna} are particularly suited for modified gravity studies since the analysis makes no {\it a priori} assumptions about the mass distribution, and it is therefore less model dependent. If one constructs first a mass model for the cluster, then one must postulate a theory of gravity to compute the lensing signal associated with that mass distribution. The lensing convergence maps obtained from such an approach could therefore be biased towards the assumed theory of gravity, which would prevent a direct comparison with other models. We have also pointed out that the analysis of Ref.~\cite{Merten:2014wna} is, however, not completely model independent, as it assumes a fiducial $\lcdm$ background model ($\Omega_{m0} = 0.27$) to compute angular diameter distances. In Secs.~\ref{sec:method}, we explained that this extra model dependency can nevertheless be taken into account by applying a correction factor, Eq.~(\ref{eq:factor}), to the convergence profiles predicted by models with different cosmological backgrounds. This correction factor holds under the approximation that all background source galaxies lie at the same redshift, which as we argued in Sec.~\ref{sec:validity}, turns out to be a good approximation with negligible impact on our conclusions.

Our main results can be summarized as follows:

\setlength{\leftmargini}{10pt}\begin{itemize}
\item Although $\Omega_{m0}$ is a parameter that enters the calculation of the lensing convergence, its impact is very small compared to the size of the effects of $M_{200}$ and $c_{200}$ (cf.~Fig.~\ref{fig:varyparams}). This means that assuming a particular value for $\Omega_{m0}$ does not introduce any significant biases in the cluster mass and concentration estimates. We have shown this explicitly for $\lcdm$ by simultaneously varying $\Omega_{m0}$, $M_{200}$ and $c_{200}$, and found barely any difference from the constraints on the cluster parameters obtained when the cosmological matter density is fixed at $\Omega_{m0} = 0.27$.

\item {The $M_{200}$ and $c_{200}$ values obtained for the CLASH clusters using GR, Cubic Galileon and Nonlocal gravity agree to better than $5\%$, which is much smaller than the $\sim 50\%-80\%$ precision allowed by the data at the $1\sigma$ level (cf.~Table \ref{table:bfmc})}. In the case of the Galileon model, this is because the screening mechanism suppresses the modifications to gravity very efficiently on the scales probed by the lensing data, $R \lesssim R_{200}$. In the case of the Nonlocal gravity model, there are no systematic shifts in the values of $M_{200}$ and $c_{200}$ relative to those in $\lcdm$ because the boost in the gravitational strength is not strong enough at the redshift of the CLASH clusters, $z \sim 0.2 - 0.9$.

\item The practically unmodified lensing masses in the Galileon model have interesting implications for tests of gravity that are designed to detect differences between lensing and dynamical mass estimates \cite{2012PhRvL.109e1301L, Lam:2013kma, Zu:2013joa}. These tests were first put forward in the context of models like $f(R)$ and DGP that modify the dynamical potential (probed by, e.g., infalling galaxies outside $R_{200}$), but not the lensing potential. Our results show that, although the Galileon model also modifies the lensing potential, this does not translate into modified lensing masses because of the screening. However, outside $R_{200}$, the force profile of the CLASH clusters in the Galileon model can be $10-40\%$ higher than in $\lcdm$ (cf.~Fig.~\ref{fig:focus}), which can affect the velocity distribution of infalling galaxies, and hence, the dynamical mass estimates. The picture is therefore qualitatively similar to that of $f(R)$ and DGP gravity, and as a result, the techniques of Refs.~\cite{2012PhRvL.109e1301L, Lam:2013kma, Zu:2013joa} can also be applied to models like the Galileon model studied here.

\end{itemize}

The existence of screening mechanisms in modified gravity models (such as the Galileon, which leads to practically unmodified lensing masses) motivates research into the lensing effects associated with cosmic voids. There, the density is low, and as a result, the fifth force effects are manifested more prominently. The lensing signal associated with voids has been detected recently in Refs.~\cite{Melchior:2013gxd, Clampitt:2014gpa} by stacking voids found in the galaxy distribution of Sloan Digital Sky Survey catalogues (see also Ref.~\cite{2013ApJ...762L..20K} for an earlier forecast study). In the context of modified gravity, Ref.~\cite{Cai:2014fma} showed that voids found in simulations are on average emptier in $f(R)$ gravity than in $\lcdm$. This happens because the enhanced gravity facilitates the pile up of matter in the surrounding walls and filaments, leaving less mass inside the void. This translates into a stronger signature in the lensing signal from voids. By the same reasons, the expectation is that voids should also be emptier in the Cubic Galileon and Nonlocal gravity models. However, in these models, one has also the effects of the modified lensing potential, which should amplify the size of an eventual signature for modified gravity. This suggests that lensing by voids could become a very powerful tool to test the law of gravity outside of the Solar System. Such an investigation is the subject of ongoing work \cite{meetal}.

\begin{acknowledgments}

We thank Mathilde Jauzac, Richard Massey and Matthieu Schaller for useful comments and discussions. We also thank Lydia Heck for invaluable numerical support. This work was supported by the Science and Technology Facilities Council [grant number ST/L00075X/1]. This work used the DiRAC Data Centric system at Durham University, operated by the Institute for Computational Cosmology on behalf of the STFC DiRAC HPC Facility (www.dirac.ac.uk). This equipment was funded by BIS National E-infrastructure capital grant ST/K00042X/1, STFC capital grant ST/H008519/1, and STFC DiRAC Operations grant ST/K003267/1 and Durham University. DiRAC is part of the National E-Infrastructure. AB is supported by FCT-Portugal through grant SFRH/BD/75791/2011. EJ is supported by Fermi Research Alliance, LLC under the U.S. Department of Energy under contract No. DE-AC02-07CH11359. LK thanks the University of Texas at Dallas for support. The research leading to these results has received funding from the People Programme (Marie Curie Actions) of the European Union's Seventh Framework Programme (FP7/2007-2013) under REA grant agreement number 627288. The research leading to these results has received funding from the European Research Council under the European Union's Seventh Framework Programme (FP/2007-2013) / ERC Grant NuMass Agreement n. [617143]. This work has been partially supported by the European Union FP7  ITN INVISIBLES (Marie Curie Actions, PITN- GA-2011- 289442) and STFC.

\end{acknowledgments}

\appendix

\section{Cluster redshifts and best-fitting parameters in $\lcdm$, Galileon and Nonlocal gravity}

\begin{table*}
\caption{CLASH cluster redshifts, $z_d$, and the effective source redshift used in the fitting, $z_s^{\rm eff}$. The values of $z_s^{\rm eff}$ are those quoted in Table 4 of Ref.~\cite{Merten:2014wna}, which correspond to the weak lensing source redshifts coming from the ground-based observations. CLJ1226+3332 is an exception since the ground based data for this cluster is not of sufficient quality (see Sec.~4.1 of Ref.~\cite{Merten:2014wna}). For this cluster we used the value of $z_s^{\rm eff}$ associated with the space-based observations. The table shows also the values of $D_d\ \left[{\rm Mpc}/h\right]$ and $\Sigma_c\ \left[10^{15} M_{\odot}h/{\rm Mpc}^2\right]$ for $\lcdm$, Galileon and Nonlocal gravity models.}
\begin{tabular}{@{}lccccccccccc}
\hline
\hline
\rule{0pt}{3ex}  Cluster name  & \ \ $z_d$ \ \  \ \  & $z^{\rm eff}_{s}$ \ \  \ \ &  $D_d^{\lcdm}$ \ \  \ \ & $\Sigma_c^{\lcdm}$ \ \  \ \ &  $D_d^{\rm Galileon}$ \ \  \ \ & $\Sigma_c^{\rm Galileon}$ \ \  \ \ &  $D_d^{\rm Nonlocal}$ \ \  \ \ & $\Sigma_c^{\rm Nonlocal}$\ \  &\ \ 
\\
\hline
\rule{0pt}{3ex}  Abell 383                   &$0.188 $  &$1.16$&  $455.62$&  $4.60$&  $462.88$&  $4.48$&  $461.47$&  $4.52$&    \\ 
\rule{0pt}{3ex}  Abell 2261                 &$0.225$   &$0.89$&  $524.38$&  $4.52$&  $534.31$&  $4.37$&  $532.09$&  $4.43$&        \\
\rule{0pt}{3ex}  MACSJ1206-08       &$0.439$   &$1.13$&    $827.59$&  $3.74$&  $856.42$&  $3.54$&  $846.15$&  $3.64$&       \\
\rule{0pt}{3ex}  RXJ1347-1145        &$0.451$   &$1.17$&    $840.69$&  $3.67$&  $870.65$&  $3.67$&  $859.80$&  $3.58$&  \\
\rule{0pt}{3ex}  MACSJ0329-02       &$0.450$   &$1.18$&    $839.62$&  $3.65$&  $869.48$&  $3.65$&  $858.67$&  $3.56$&  \\
\rule{0pt}{3ex}  MS2137-2353         &$0.313$    &$1.23$&    $666.70$&  $3.67$&  $683.94$&  $3.65$&  $679.02$&  $3.58$&  \\
\rule{0pt}{3ex}  MACSJ0744+39      &$0.686$   &$1.41$&    $1037.60$& $3.84$&  $1087.94$& $3.84$& $1065.27$&  $3.73$& \\
\rule{0pt}{3ex}  MACSJ1115+0129 &$0.352$   &$1.15$&    $721.39$&  $3.68$&  $742.14$&  $3.68$&  $735.72$&  $3.59$&  \\
\rule{0pt}{3ex}  Abell 611                  &$0.288$   &$1.13$&    $629.07$&  $3.86$&  $644.13$&  $3.86$&  $640.08$&  $3.77$&  \\
\rule{0pt}{3ex}  RXJ1532.8+3021   &$0.363$   &$1.15$&     $735.97$&  $3.67$&  $757.73$&  $3.67$&  $750.86$&  $3.58$&   \\
\rule{0pt}{3ex}  MACSJ1720+3536 &$0.391$  &$1.13$&     $771.54$&  $3.69$&  $795.89$&  $3.69$&  $787.81$&  $3.60$&  \\
\rule{0pt}{3ex}  RXJ2129+0005      &$0.234$   &$1.12$&     $540.25$&  $4.12$&  $550.88$&  $4.12$&  $548.43$&  $4.05$&  \\
\rule{0pt}{3ex}  MACSJ1931-26      &$0.352$   &$1.16$&     $721.39$&  $4.10$&  $742.14$&  $4.10$&  $735.72$&  $3.99$&  \\
\rule{0pt}{3ex}  Abell 209                 &$0.206$    &$0.93$&    $489.79$&  $4.60$&  $498.31$&  $4.60$&  $496.53$&  $4.52$&  \\
\rule{0pt}{3ex}  RXCJ2248-4431    &$0.348$    &$0.94$&     $715.99$&  $3.84$&  $736.38$&  $4.60$&  $730.12$&  $3.74$&  \\
\rule{0pt}{3ex}  MACSJ0429-02      &$0.399$   &$1.25$&     $781.30$&  $3.52$&  $806.39$&  $3.52$&  $797.97$&  $3.43$&   \\
\rule{0pt}{3ex}  MACSJ1423+24     &$0.5454$ &$0.98$&     $932.09$&  $4.70$&  $970.66$&  $4.70$&  $955.11$&  $4.56$&  \\
\rule{0pt}{3ex}  CLJ1226+3332       &$0.890$   &$1.66$&     $1140.34$& $4.12$& $1203.90$& $4.12$&  $1172.32$&  $4.02$&  \\
\rule{0pt}{3ex}  MACSJ1311-03      &$0.494$   &$1.07$&     $884.88$&  $4.03$&  $918.83$&  $4.03$&  $905.86$&  $3.92$&    \\
\hline
\hline
\end{tabular}
\label{table:d}
\end{table*} 

\begin{table*}
\caption{Best-fitting $M_{200}\ \left[10^{15}M_{\odot}/h\right]$ and $c_{200}$ values for all the CLASH clusters in the $\lcdm$, Galileon and Nonlocal gravity models. The errors quoted correspond to the marginalized 1-$\sigma$ limits.}
\begin{tabular}{@{}lccccccccccc}
\hline
\hline
\rule{0pt}{3ex}  Cluster name  & \ \ $M_{200}^{\lcdm}$ \ \  \ \  & $c_{200}^{\lcdm}$ \ \  \ \ &  $M_{200}^{\rm Galileon}$ \ \  \ \ & $c_{200}^{\rm Galileon}$ \ \  \ \ &  $M_{200}^{\rm Nonlocal}$ \ \  \ \ & $c_{200}^{\rm Nonlocal}$ &\ \ 
\\
\hline
\rule{0pt}{3ex}  Abell 383                        & $ 1.01 \pm 0.22 $& $ 4.54 \pm 1.17 $& $ 0.98 \pm 0.21 $ & $ 4.68 \pm 1.17 $ &  $ 0.98 \pm 0.21 $&  $ 4.46 \pm 1.13 $&   \\ 
\rule{0pt}{3ex}  Abell 2261                     & $ 1.10 \pm 0.35 $& $ 2.91 \pm 1.03 $&  $ 1.07 \pm 0.32 $& $ 2.98 \pm 1.06 $ &  $ 1.07 \pm 0.32 $&  $ 2.84 \pm 1.03 $&   \\ 
\rule{0pt}{3ex}  MACSJ1206-08            & $ 0.83 \pm 0.16 $& $ 4.18 \pm 1.06 $&  $ 0.80 \pm 0.16 $& $ 4.32 \pm 1.10 $ &  $ 0.83 \pm 0.16 $&  $ 4.18 \pm 1.06 $&   \\ 
\rule{0pt}{3ex}  RXJ1347-1145             & $ 1.07 \pm 0.21 $& $ 3.19 \pm 0.85 $&  $ 1.04 \pm 0.21 $& $ 3.33 \pm 0.88 $ &  $ 1.07 \pm 0.21 $&  $ 3.19 \pm 0.85 $&   \\ 
\rule{0pt}{3ex}  MACSJ0329-02            & $ 0.72 \pm 0.16 $& $ 3.69 \pm 0.88 $&  $ 0.69 \pm 0.16 $& $ 3.83 \pm 0.92 $ &  $ 0.69 \pm 0.16 $&  $ 3.69 \pm 0.88 $&   \\ 
\rule{0pt}{3ex}  MS2137-2353               & $ 1.33 \pm 0.35 $& $ 2.70 \pm 0.64 $&  $ 1.27 \pm 0.34 $& $ 2.84 \pm 0.64 $ &  $ 1.30 \pm 0.34 $&  $ 2.70 \pm 0.64 $&   \\ 
\rule{0pt}{3ex}  MACSJ0744+39           & $ 0.69 \pm 0.13 $& $ 3.97 \pm 0.78 $&  $ 0.69 \pm 0.13 $& $ 4.11 \pm 0.81 $ &  $ 0.69 \pm 0.13 $&  $ 3.97 \pm 0.78 $&   \\ 
\rule{0pt}{3ex}  MACSJ1115+0129      & $ 0.89 \pm 0.18 $& $ 2.20 \pm 0.39 $&  $ 0.86 \pm 0.16 $& $ 2.27 \pm 0.42 $ &  $ 0.86 \pm 0.16 $&  $ 2.20 \pm 0.39 $&   \\ 
\rule{0pt}{3ex}  Abell 611                       & $ 0.83 \pm 0.22 $& $ 3.26 \pm 0.64 $&  $ 0.80 \pm 0.21 $& $ 3.33 \pm 0.67 $ &  $ 0.80 \pm 0.21 $&  $ 3.26 \pm 0.64 $&   \\ 
\rule{0pt}{3ex}  RXJ1532.8+3021        & $ 0.54 \pm 0.12 $& $ 2.84 \pm 0.81 $&  $ 0.51 \pm 0.12 $& $ 2.98 \pm 0.81 $ &  $ 0.51 \pm 0.12 $&  $ 2.84 \pm 0.81 $&   \\ 
\rule{0pt}{3ex}  MACSJ1720+3536     & $ 0.72 \pm 0.15 $& $ 4.11 \pm 1.10 $&  $ 0.72 \pm 0.15 $& $ 4.25 \pm 1.13 $ &  $ 0.72 \pm 0.15 $&  $ 4.04 \pm 1.10 $&   \\ 
\rule{0pt}{3ex}  RXJ2129+0005           & $ 0.60 \pm 0.13 $& $ 4.18 \pm 0.85 $&  $ 0.57 \pm 0.13 $&  $ 4.32 \pm 0.88 $&  $ 0.57 \pm 0.13 $&  $ 4.11 \pm 0.85 $&   \\ 
\rule{0pt}{3ex}  MACSJ1931-26           & $ 0.48 \pm 0.16 $& $ 3.19 \pm 1.20 $&  $ 0.48 \pm 0.15 $& $ 3.33 \pm 1.24 $ &  $ 0.48 \pm 0.16 $&  $ 3.19 \pm 1.20 $&   \\ 
\rule{0pt}{3ex}  Abell 209                      & $ 0.92 \pm 0.21 $& $ 3.12 \pm 0.74 $&  $ 0.86 \pm 0.19 $&  $ 3.19 \pm 0.74 $&  $ 0.86 \pm 0.19 $&  $ 3.05 \pm 0.74 $&   \\ 
\rule{0pt}{3ex}  RXCJ2248-4431         & $ 1.10 \pm 0.31 $& $ 2.98 \pm 1.24 $&  $ 1.07 \pm 0.29 $&  $ 3.12 \pm 1.27 $&  $ 1.07 \pm 0.31 $&  $ 2.98 \pm 1.24 $&   \\ 
\rule{0pt}{3ex}  MACSJ0429-02          & $ 0.77 \pm 0.19 $& $ 3.05 \pm 1.52 $&  $ 0.74 \pm 0.18 $&  $ 3.26 \pm 1.59 $&  $ 0.74 \pm 0.18 $&  $ 3.05 \pm 1.52 $&   \\ 
\rule{0pt}{3ex}  MACSJ1423+24         & $ 0.54 \pm 0.13 $& $ 4.61 \pm 1.27 $&  $ 0.54 \pm 0.13 $&  $ 4.82 \pm 1.31 $&  $ 0.54 \pm 0.13 $&  $ 4.61 \pm 1.27 $&   \\ 
\rule{0pt}{3ex}  CLJ1226+3332           & $ 1.51 \pm 0.32 $& $ 3.90 \pm 0.71 $&  $ 1.56 \pm 0.34 $&  $ 3.97 \pm 0.71 $&  $ 1.54 \pm 0.32 $&  $ 3.90 \pm 0.71 $&   \\ 
\rule{0pt}{3ex}  MACSJ1311-03          & $ 0.45 \pm 0.07 $& $ 4.32 \pm 0.64 $&  $ 0.42 \pm 0.07 $&  $ 4.46 \pm 0.67 $&  $ 0.45 \pm 0.07 $&  $ 4.25 \pm 0.64 $&   \\ 
\hline
\hline
\end{tabular}
\label{table:bfmc}
\end{table*} 

\bibliography{clusterlensing.bib}

\end{document}